\documentclass[aps,prx,twocolumn,superscriptaddress,a4paper,english,longbibliography]{revtex4-2}
\usepackage{times}
\usepackage{color}
\usepackage[colorlinks=true,urlcolor=blue,citecolor=blue]{hyperref}
\usepackage{url}
\usepackage{breakurl}
\usepackage{soul}
\usepackage{graphicx}
\usepackage{amsmath}
\usepackage{subfigure}
\usepackage{xcolor}
\usepackage{amssymb}
\usepackage{latexsym}
\usepackage{hyperref}
\DeclareGraphicsExtensions{.png,.eps}
\usepackage{float}
\usepackage{appendix}
\usepackage{etoolbox}

\newcommand {\R}{\textcolor {black}}

\usepackage{xcolor}
\usepackage[urlcolor=blue]{hyperref}      
\hypersetup{
	colorlinks = true,                    
	citecolor = {blue},
	linkcolor = {purple},
}

\begin{document}	
    \title{\R{Non-parametric Semi-Supervised Learning in Many-body Hilbert Space with Rescaled Logarithmic Fidelity}}
	
\author{Wei-Ming Li}
\affiliation{Department of Physics, Capital Normal University, Beijing 100048, China}

\author{Shi-Ju Ran}
\email[Corresponding author. Email: ] {sjran@cnu.edu.cn}
\affiliation{Department of Physics, Capital Normal University, Beijing 100048, China}

\begin{abstract}
	In quantum and quantum-inspired machine learning, the very first step is to embed the data in quantum space known as Hilbert space. Developing quantum kernel function (QKF), which defines the distances among the samples in the Hilbert space, belongs to the fundamental topics for machine learning. In this work, we propose the rescaled logarithmic fidelity (RLF) and non-parametric semi-supervised learning in the quantum space, which we name as RLF-NSSL. The rescaling takes advantage of the non-linearity of the kernel to tune the mutual distances of samples in the Hilbert space, and meanwhile avoids the exponentially-small fidelities between quantum many-qubit states. Being non-parametric excludes the possible effects from the variational parameters, and evidently demonstrates the advantages from the space itself. We compare RLF-NSSL with several well-known non-parametric algorithms including naive Bayes classifiers, $k$-nearest neighbors, and spectral clustering. Our method exhibits \R{better} accuracy particularly for the unsupervised case with no labeled samples and the few-shot cases with small numbers of labeled samples. With the visualizations by t-stochastic neighbor embedding, our results imply that the machine learning in the Hilbert space complies with the principles of maximal coding rate reduction, where the low-dimensional data exhibit within-class compressibility, between-class discrimination, and overall diversity. Our proposals can be applied to other quantum and quantum-inspired machine learning, including the methods using the parametric models such as tensor networks, quantum circuits, and quantum neural networks.
\end{abstract}
\maketitle

\section{Introduction}

In machine learning and other relevant fields, kernel
\cite{shawe2004kernel,hofmann2008kernel}
is defined as a special function to characterize the similarity (or distance) between any two samples after mapping them to a high-dimensional space. For the quantum and quantum-inspired machine learning (QML in short), an essential step to process classical data is to embed the data (e.g., images, texts, and etc.) to the quantum space known as Hilbert space \cite{biamonte2017quantum, schuld2019quantum, havlivcek2019supervised, lloyd2020quantum, schuld2021supervised}. The quantum kernel function (QKF) that characterizes the distributions in the Hilbert space
\cite{wiebe2012quantum, lloyd2014quantum, stoudenmire2016supervised, schuld2016prediction, benedetti2017quantum,schuld2017implementing, biamonte2017quantum, schuld2019quantum, kerenidis2019q,havlivcek2019supervised, zhao2019quantum, lloyd2020quantum, larose2020robust,huang2021power, park2020theory, schuld2021supervised}
is usually a critical factor for the performance of a QML scheme. 

Among the existing mappings from data to the quantum state representations, a widely recognized example is known as the quantum feature map (see, e.g., \cite{stoudenmire2016supervised, han2018unsupervised, liu2019machine, sun2020generative}). It maps each feature to the state of one qubit, and each sample to an $M$-qubit product state (with $M$ the number of features). Such a quantum feature map brings interpretability from the perspective of quantum probabilities, and have succeeded in the supervised \cite{stoudenmire2016supervised, liu2019machine, sun2020generative} and unsupervised learning \cite{han2018unsupervised, ran2020tensor} algorithms as well as in the QML experiments \cite{wang2020quantum}. It is unexplored to use such a quantum feature map for semi-supervised learning.

In quantum information and computation \cite{nielsen2002quantum}, fidelity serves as a fundamental quantity to characterize the similarity of two quantum states, and has been applied in tomography \cite{d2001quantum}, verification \cite{buhrman2006quantum}, and detection of quantum phase transitions \cite{zhou2008ground, abasto2008fidelity, schwandt2009quantum, quan2009quantum, zhao2009singularities, xiong2009reduced}. One drawback of fidelity for the many-qubit states is that it usually decreases exponentially with the number of qubits $M$, which is known as the ``orthogonal catastrophe''. Instability or overflow of the precisions may occur for large $M$. One way to avoid the ``orthogonal catastrophe'' is to use the logarithmic fidelity (for instance, \cite{zhou2008ground, ran2020encoding, yang2021visualizing}). However, it is unclear how the mutual distances of the samples or the data structure will be altered by taking logarithm on the fidelity.

In this work, we propose the rescaled logarithmic fidelity (RLF) as a tunable QKF. To show its validity, we implement non-parametric semi-supervised learning in the Hilbert space based on RLF, which we name as RLF-NSSL. Being non-parametric, we can exclude the possible effects from the variational parameters and focus on the space and kernel. \R{Note for the parametrized models, say neural networks, the performances are mainly determined by their architecture and parameter complexities, such as the arrangements of different types of layers and the numbers of variational parameters therein.} In the RLF-NSSL, a give sample is classified by comparing the RLF's between this sample and the clusters that are formed by labeled and pseudo-labeled samples. \R{A strategy for pseudo-labeling is proposed}. RLF-NSSL achieves \R{better} accuracy comparing with several established non-parametric methods such as naive Bayes classifiers, $k$-nearest neighbors, and spectral clustering. Particularly for the unsupervised or few-shot cases where the choice of kernel is crucial, the high performance of our method indicates the validity of RLF for QML.

The clusters formed by the samples with labels and pseudo-labels also define a mapping from the original space to a low-dimensional effective space. With the visualization by t-SNE \cite{van2008visualizing}, we show that the low-dimensional data exhibit within-class compressibility, between-class discrimination, and overall diversity. It implies the machine learning in the Hilbert space also comply with the principles of maximal coding rate reduction \cite{ma2007segmentation, yu2020learning}. We expect that these findings, including the RLF \R{and the pseudo-labeling strategy}, would generally benefit the 
QML using the parametric models, such as tensor networks \cite{socher2013reasoning,  stoudenmire2016supervised, han2018unsupervised, cheng2018information, chen2018equivalence, cheng2019tree, liu2019machine, huggins2019towards, schroder2019tensor, efthymiou2019tensornetwork, sun2020generative, sun2020tangent, guo2020tensor, ran2020tensor, cheng2021supervised, reyes2021multi}, parametric quantum circuits 
\cite{zhu2019training, benedetti2019generative,benedetti2019parameterized,chen2020variational,du2020expressive,cao2020cost,du2020expressive, xin2021experimental, cincio2021machine}, and quantum neural networks \cite{farhi2018classification, mcclean2018barren, cong2019quantum, killoran2019continuous, mari2020transfer, beer2020training, shen2020information}. \R{The connections to the maximal coding rate reduction shows the possibility of understanding the QML from the perspective of coding rate.}

\section{Hilbert space and rescaled logarithmic fidelity}

Given a sample that we assume to be a $M$-component vector $\mathbf{x} = \{x_1, x_2, \ldots, x_M\}$ with $0 \leq x_m \leq 1$, the feature map (see, e.g., \cite{stoudenmire2016supervised, han2018unsupervised, liu2019machine, sun2020generative}) to encode it to a $M$-qubit product states is written as 
\begin{eqnarray}
 |\phi \rangle = \prod_{\otimes m=1}^M \left[\cos(\frac{x_m\pi}{2}) |0_m\rangle + \sin(\frac{x_m\pi}{2}) |1_m\rangle \right].
 \label{eq-featuremap}
\end{eqnarray}
Here, $|0_m\rangle$ and $|1_m\rangle$ form a set of orthonormal basis for the $m$-th qubit, which satisfy $\langle a_m|b_m \rangle = \delta_{ab}$. 

In quantum information, the quantity to characterize the similarity between two states $|\phi^1 \rangle$ and $|\phi^2 \rangle$ is the fidelity $f$ defined as the absolute value of their inner product $f (|\phi^1 \rangle, |\phi^2 \rangle) = |\langle \phi^1|\phi^2 \rangle|$. As each state is normalized, we have $0 \leq f \leq 1$. With $f=0$, the two states are orthogonal to each other and have the smallest similarity. With $f=1$, the states satisfy $|\phi^1 \rangle = e^{i\alpha} |\phi^2 \rangle$ with $\alpha$ a universal phase factor. In this case, $|\phi^1 \rangle$ and $|\phi^2 \rangle$ can be deemed as a same state (meaning zero distance). 

The fidelity with the feature map in Eq. (\ref{eq-featuremap}) results in a QKF to characterize the similarity between two samples $\mathbf{x}^1$ and $\mathbf{x}^2$, which reads
\begin{eqnarray}
	f(\mathbf{x}^1, \mathbf{x}^2) = \prod_{m=1}^M \left| \cos\left[\frac{\pi}{2} (x^1_m - x^2_m) \right] \right|.
	\label{eq-kernel}
\end{eqnarray}
In other words, the similarity between $\mathbf{x}^1$ and $\mathbf{x}^2$ is characterized by the fidelity $f$ between the two product states obtained by implementing the feature map on these two samples.

Eq. (\ref{eq-kernel}) shows that the fidelity is the product of $M$ non-negative numbers $\left| \cos\left[\frac{\pi}{2} (x^1_m - x^2_m) \right] \right| \leq 1$ (the equality holds when the corresponding feature takes the same value in the two samples, i.e.,  $x^1_m = x^2_m$). Consequently, $f(\mathbf{x}^1, \mathbf{x}^2)$ decreases exponentially with the number of pixels that take different values in $\mathbf{x}^1$ and $\mathbf{x}^2$. Taking MNIST dataset as an example, there are usually $O(10^2)$ such pixels. Then $f$ will be extremely small, meaning that the states from any two of the samples are almost orthogonal to each other. This is known as the ``orthogonal catastrophe'', where instability or precision overflow may occur.

One way to resolve this problem is to use the logarithmic fidelity (for instance, \cite{zhou2008ground, ran2020encoding, yang2021visualizing})
\begin{eqnarray}
	F(\mathbf{x^1}, \mathbf{x^2}) &=& \log_{10} f(\mathbf{x^1}, \mathbf{x^2}) \nonumber \\
	&=& \sum_{m=1}^M \log \left\{ \left| \cos\left[\frac{\pi}{2} (x^1_m - x^2_m) \right] \right|  + \varepsilon\right\}
	\label{eq-NLF}
\end{eqnarray}
with $\varepsilon$ a small positive constant to avoid $\log 0$. $F$ is a non-positive scalar that also characterizes the similarity between the given states. Though $F$ changes monotonously with $f$, the mutual distances among the samples obtained by these two kernels are definitely different. For instance, we might have $\sum_n F(\mathbf{x^1}, \mathbf{x^n}) < \sum_n F(\mathbf{x^2}, \mathbf{x^n})$ while $\sum_n f(\mathbf{x^1}, \mathbf{x^n}) > \sum_n f(\mathbf{x^2}, \mathbf{x^n})$,  due to the nonlinearity of the logarithmic function.

\section{Rescaled logarithmic fidelity and classification scheme}
\label{sec3}

In this work, we take advantage of the nonlinearity and define the \textit{rescaled logarithmic fidelity} (RLF) as
\begin{eqnarray}
\tilde{f}_{\beta} (\mathbf{x}^1, \mathbf{x}^2) =  \beta^{ F(\mathbf{x^1}, \mathbf{x^2})}= \beta^{\log _{10} f(\mathbf{x^1}, \mathbf{x^2})}
\label{eq-RLF}
\end{eqnarray}
with $\beta$ a tunable parameter that we dub as the \textit{rescaling factor}. In particular for $\beta=10$, the RLF becomes the fidelity, i.e., $\tilde{f}_{10} (\mathbf{x}^1, \mathbf{x}^2) = f(\mathbf{x}^1, \mathbf{x}^2)$.

\begin{figure*}[tbp]
	\includegraphics[width=0.9\linewidth]{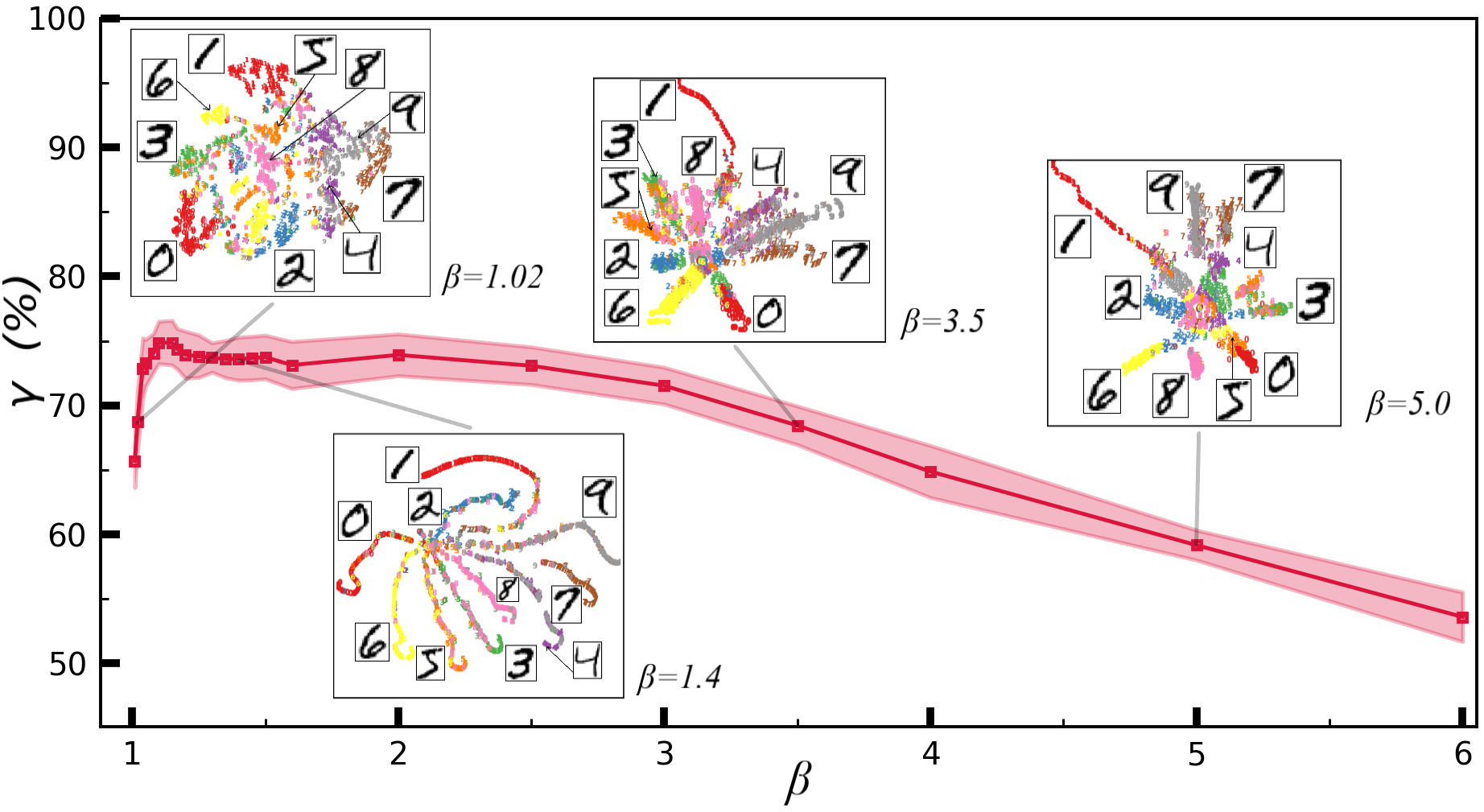}
	\caption{(Color online) The $\beta$ dependence of the testing accuracy $\gamma$ of the testing samples for the ten classes in the MNIST dataset, where $\varepsilon_t$ is evaluated by randomly taking N=10 training samples from each classes. We take the average of $\varepsilon_t$ by implement the simulations for 20 times, and the variances are indicated by the shadowed area. By t-SNE, the insets show the visualized distributions of 2000 effective vectors $\{\tilde{\mathbf{y}}\}$ [Eq. (\ref{eq-aveRLF})] that are randomly taken from the testing samples.}
	\label{fig-base}
\end{figure*}

With certain labeled training samples $\{\mathbf{x}^l\}$ from $P$ classes, an unlabeled sample $\mathbf{y}$ can be classified in a supervised learning process. First, we transform $\mathbf{y}$ to a $P$-dimensional effective vector $\tilde{\mathbf{y}}$, where its $p$-th element is the average RLF with the training samples from the $p$-th class
\begin{eqnarray}
	\tilde{y}_p = \frac{1}{N_p} \sum_{\mathbf{x}^l \in \text{class-$p$}} \tilde{f}_{\beta}(\mathbf{x}^l, \mathbf{y}),
	\label{eq-aveRLF}
\end{eqnarray}
with $N_p$ the number of the labeled samples that belong to the $p$-th class. We call the labeled samples in a same class as a \textit{cluster}. The clusters define a dimensionality reduction map given by Eq. (\ref{eq-aveRLF}) from the original feature space to a $P$-dimensional space. In practice, we take $N_{p}=N$ as a same number for all $p$. The classification of $\mathbf{y}$ is then indicated by the largest element of $\tilde{\mathbf{y}}$ as
\begin{eqnarray}
	c(\mathbf{y}) = \arg\max_p(\tilde{y}_p).
	\label{eq-pred}
\end{eqnarray}
One can see that except certain hyper-parameters such as the rescaling factor $\beta$ and the number of labeled samples, the above method contains no variational parameters, thus is dubbed as non-parametric supervised learning with RLF (RLF-NSL in short).

\R{Classically, RLF can be easily calculated. Therefore, the classification algorithms based on RLF can be regarded as the quantum-inspired machine learning schemes running on classical computers. Considering to run such algorithms on the quantum platforms, the main challenge is the estimation of Eq. (\ref{eq-aveRLF}) in order to obtain the similarity between a given sample and the clusters. It requires to estimate the rescaled logarithmic fidelity $\tilde{f}_{\beta}$ and calculate the summation over the samples in the cluster. In our cases, estimating $\tilde{f}_{\beta}$ is much easier than estimating the fidelity or implementing the full-state tomography for arbitrary states, since it is essentially the fidelity between two product states. Quantum acceleration over classical computation is unlikely in calculating such a fidelity, however, it is possible to gain quantum acceleration by parallelly computing the summations over the samples. This requires to design the corresponding quantum circuit regarding RLF, which is an open issue for the future investigations.}

To demonstrate how $\beta$ affects the classification accuracy, we choose the MNIST dataset \cite{mnistweb} as an example, and randomly take $N=10$ labeled samples from each class of the training set. \R{The MNIST dataset contains the grey-scale images of hand-written digits, where the resolution of each image is $28 \times 28$ (meaning $784$ features in each image). The images are divided into two sets with $60000$ images as the training samples and $10000$ as the testing samples.} We obtain the effective vectors $\{\tilde{\mathbf{y}}\}$ of all testing samples using Eq. (\ref{eq-aveRLF}), and calculate the classification using Eq. (\ref{eq-pred}). The testing accuracy $\gamma$ is calculated as the number of the correctly classified testing samples divided by the total number. Fig. \ref{fig-base} shows the $\gamma$ when the number of the labeled samples in each class is small (few-shot learning with $N=10$). We show the average of $\gamma$ by implementing the simulations for 20 times, and the variances are illustrated by the shadowed areas. All the variances in this paper are obtained a similar way. One can see $\gamma$ firstly rises and then drops by increasing $\beta$, and reaches the maximum around $1.2 < \beta < 2$. Note the $\beta$ that gives the maximal $\gamma$ slightly changes with different $N$. 

In the insets, we randomly take $200$ testing samples from each class, and reduce the dimension of the effective vectors $\{\tilde{\mathbf{y}}\}$ from $10$ to $2$ by t-SNE \cite{van2008visualizing}, in order to visualize the distribution of the testing samples. \R{The t-SNE is a non-linear dimensionally reduction method. It maps the given samples to a lower-dimensional space by reducing the number of features. The reduction is optimal in the sense that the mutual distances (or similarities) among the samples in the lower-dimensional space should be close to those in the original space.}  By eyes one can observe better separation for larger $\gamma$ (e.g., $\beta=1.4$) compared with those $\beta$'s giving smaller $\gamma$. More discussions are given below from the perspective of rate reduction \cite{ma2007segmentation,yu2020learning}.  We also confirm with more simulations that the fidelity (equivalently with $\beta=10$ in the RLF) gives lower accuracy with $\gamma \simeq 50\%$. Note this accuracy is also not stable since the fidelity is exponentially small.

\begin{figure}[tbp]
	\includegraphics[width=1\linewidth]{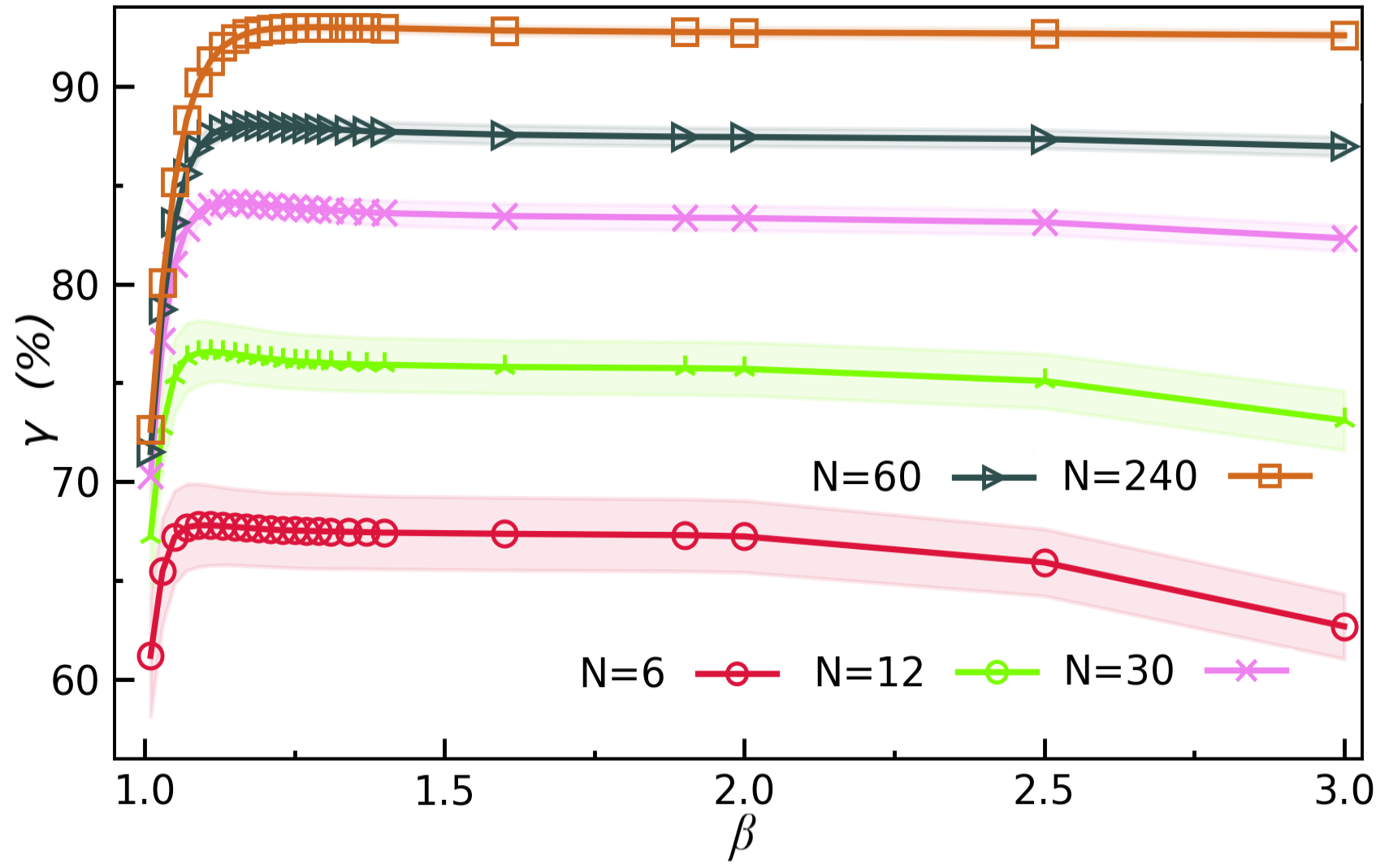}
	\caption{(Color online) Testing accuracy $\gamma$ of non-parametric supervised learning using rescaled logarithmic fidelity (RLF-NSL in short) on the MNIST dataset with different number of labeled samples $N$ in each class.}
	\label{fig-NSLbeta}
\end{figure}

Fig. \ref{fig-NSLbeta} demonstrates how the testing accuracy of RLF-NSL is affected by $\beta$ with different numbers of labeled samples $N$ in each class. In all $N$'s that varies from $6$ to $240$, $\gamma$ firstly rapidly rises and the slowly decreases with $\beta$. Approximately for $\beta \simeq 1.3$, relatively high testing accuracy is obtained in all cases.

\section{Non-parametric semi-supervised learning with pseudo-labels}
\label{sec4}

Based on RLF, we propose a non-parametric semi-supervised learning algorithm (RLF-NSSL in short). \R{Different from supervised learning where sufficiently many labeled samples are required to implement the machine learning tasks, the key of semi-supervised learning is, in short, to utilize the samples whose labels are predicted by the algorithm itself. The generated labels are called pseudo-labels. The supervised learning can be considered as a special case of semi-supervised learning with zero pseudo-labels. For the unsupervised kernel-based classifications where there is no labeled samples, pseudo-labels can be useful to implement the classification tasks in a way similar to the supervised cases. The strategy of tagging the pseudo-labels is key to the prediction accuracy. Therefore, for the unsupervised (and also the few-shot cases with a small number of labeled samples), the performance should strongly rely on the choice of kernel.} Here, we define $P$ clusters, of which each contains two parts: all the $N$ labeled training samples in this class and $\tilde{N}$ unlabeled samples that are classified to this class. The rescaling factor is taken as the optimal $\beta$ with $(N+\tilde{N})$ labeled samples in RLF-NSL. The key is how to choose the $\tilde{N}$ samples with pseudo-labels to expand the clusters.

Our strategy is to divide the unlabeled training samples into batches for classification and pseudo-labeling. The clusters are initialized by the labeled samples. Then for each batch of the unlabeled samples, we classify them by calculating the effective vectors given by Eq. (\ref{eq-aveRLF}), where the summation is over all the samples with labels and  pseudo-labels (if any). Then we add these samples to the corresponding clusters according to their classifications. The cluster is used to classify the testing set after all unlabeled training samples are classified. 

Inevitably the incorrect pseudo-labels would be introduced into the clusters, which may harm the classification accuracy. Therefore, we propose to update the samples in the clusters. To this aim, we define the confidence. For a sample $\mathbf{y}$ in the $p$-th cluster, it is defined as
\begin{eqnarray}
	\eta = \frac{\tilde{y}_{p}}{\sum_{p'=1}^{P} \tilde{y}_{p'}},
	\label{eq-confidence}
\end{eqnarray}
with $\tilde{y}_{p}$ obtained by Eq.~(\ref{eq-aveRLF}). Then in each cluster, we keep $N_{\Delta}$ pseudo-labels with the highest confidence. The rest pseudo-labels are removed, and the corresponding samples are thrown to the pool of the unlabeled samples, which are to be classified in the future iterations. 

\begin{figure}[tbp]
	\includegraphics[width=1\linewidth]{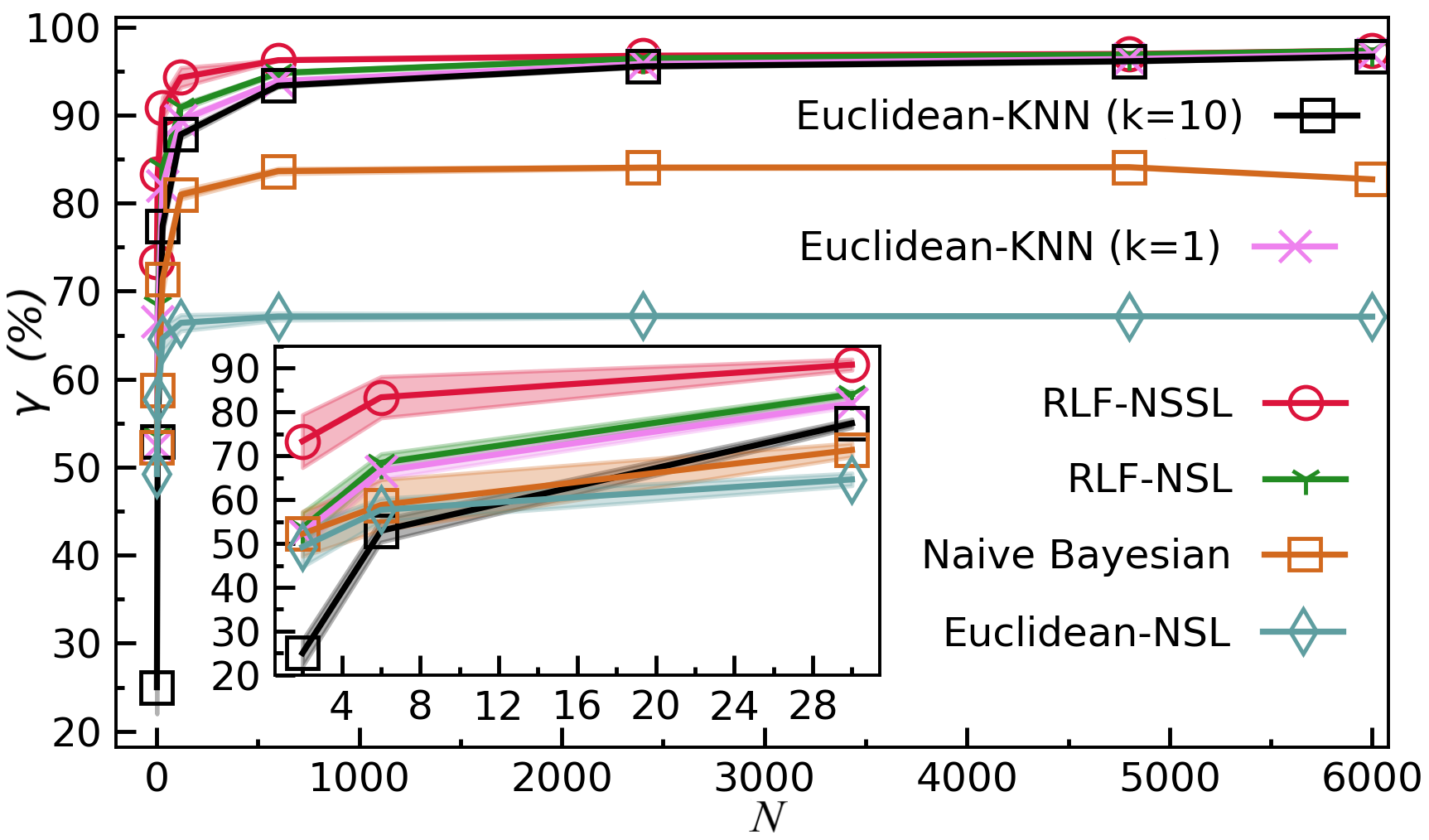}
	\caption{(Color online) Testing accuracy $\gamma$ of non-parametric semi-supervised and supervised learning using rescaled logarithmic fidelity (RLF-NSSL and RLF-NSL in short, respectively) on the MNIST dataset. Our results are compared with $k$-nearest neighbors with $k=1$ and $10$, naive Bayesian classifiers, and a baseline model by simply replacing RLF by the Euclidean distance.}
	\label{fig-supervised}
\end{figure}

Fig. \ref{fig-supervised} shows the testing accuracy $\gamma$ of the MNIST dataset with different numbers of the labeled samples $N$. The accuracy of RLF-NSL (green line) already surpasses the recognized non-parametric methods including $k$-nearest neighbors (KNN) \cite{cover1967nearest} with $k=1$ and $10$, and the naive Bayesian classifier \cite{langley1992analysis}. \R{KNN is also a kernel-based classification method. One first calculates the distances (or similarities) between the target sample and all labeled samples, and then find the $k$ labeled samples with the smallest distances. The classification of the target sample is given by finding the class that has the largest number in these $k$ samples.} The performance of RLF-NSL significantly surpasses a baseline model, where we simply replace the RLF $f_{\beta}$ in Eq. (\ref{eq-aveRLF}) by the Euclidean kernel
\begin{eqnarray}
	\R{f_E = ||\mathbf{x}^l - \mathbf{y}||.}
	\label{eq-Euclidean}
\end{eqnarray}
The RLF-NSSL achieves the highest accuracy among all the presented methods. Significant improvement is observed particularly for the few-shot learning with small $N$, as shown in the inset. Note for different $N$, we optimize $\beta$ in RLF-NSL and we fix $\beta=1.3$ in RLF-NSSL.

For the unsupervised learning, we assume there are no labeled samples from the very beginning. All samples in the clusters will be those with pseudo-labels. To start with, we randomly chose one sample to form a cluster. From all the unlabeled samples, every time we select a new sample (denoted as $\tilde{x}$) that satisfies two conditions:
\begin{eqnarray}
	\tilde{\mathbf{x}} = \text{argmax}_{\mathbf{x}} \sum_{\mathbf{x}_i \in \text{clusters}} F(\mathbf{x}, \mathbf{x}_i);\\
	F(\tilde{\mathbf{x}}, \mathbf{x}_i) < \mu \text{    for $\forall$ $\mathbf{x}_i \in$ clusters}.
\end{eqnarray}
with $\mu$ a preset small constant. Repeating the above procedure for $(P-1)$ times, we have $P$ clusters, of which each contains one sample. These samples have relatively small mutual similarities, thus are reasonable choices to initialize the clusters. We classify all the samples out of the clusters using the method explained in Sec.~\ref{sec3}. Then all samples will be added to the corresponding clusters according to the classifications. 

The next step is to use the semi-supervised learning method introduced in Sec.~\ref{sec4} to update the samples in the clusters. In specific, we remove the pseudo-labels for part of the samples in each cluster with the lowest confidence, and throw them to the pool of the unlabeled samples. We subsequently classify all the unlabeled samples and add them to the clusters correspondingly. Repeat the processes above until the clusters converge. 

\begin{table}[tbp]
	\begin{center}
		\begin{tabular*}{8.5cm}{@{\extracolsep{\fill}}lcccc}
			\hline\hline 
		       & $k$-means & spectral clustering & RLF-NSSL ($N=0$) \\ \hline
			$\gamma$ (\%)  & 56.21 & 65.46 & 72.64 \\ \hline
			Std.  & 1.83 & 0.1 & 5.43 \\ 
			\hline \hline
		\end{tabular*}
	\end{center}
	\caption{The testing accuracy $\gamma$ and the standard deviation (std.) on the MNIST dataset using $k$-means, spectral clustering, and RLF-NSSL ($N=0$, i.e., no labeled samples). We use the way proposed in \cite{munkres1957algorithms} to determine the labels of the clusters in the case of unsupervised learning. For the $k$-means, we use the randomly initialized clustering center and take 270 iteration steps. The similarity is characterized by Euclidean distance. For spectral clustering, we use the SpectralClustering function from the ``sklearn'' package in Python.}
	\label{tab}
\end{table}

Table~\ref{tab} compares the testing accuracy $\gamma$ of our RLF-NSSL with other two unsupervised methods $k$-means \cite{sarle1990algorithms, kaufman2009finding} and spectral clustering \cite{mehrotra1997elements, ng2001spectral, kamvar2003spectral}. We use the way proposed in \cite{munkres1957algorithms} to determine the labels of the clusters. For each iteration in the RLF-NSSL to update the clusters, we remove the pseudo-labels of $35 \%$ of the samples with the lowest confidence in each cluster, which are to be re-classified in the next iteration. Our RLF-NSSL achieves the highest accuracy among these three methods. RLF-NSSL exhibits relatively high standard deviation, possibly due to the large fluctuation induced by the (nearly) random initialization of the clusters. Such fluctuation can be suppressed by incorporating with a proper initialization strategy.

\begin{figure}[tbp]
	\includegraphics[width=1\linewidth]{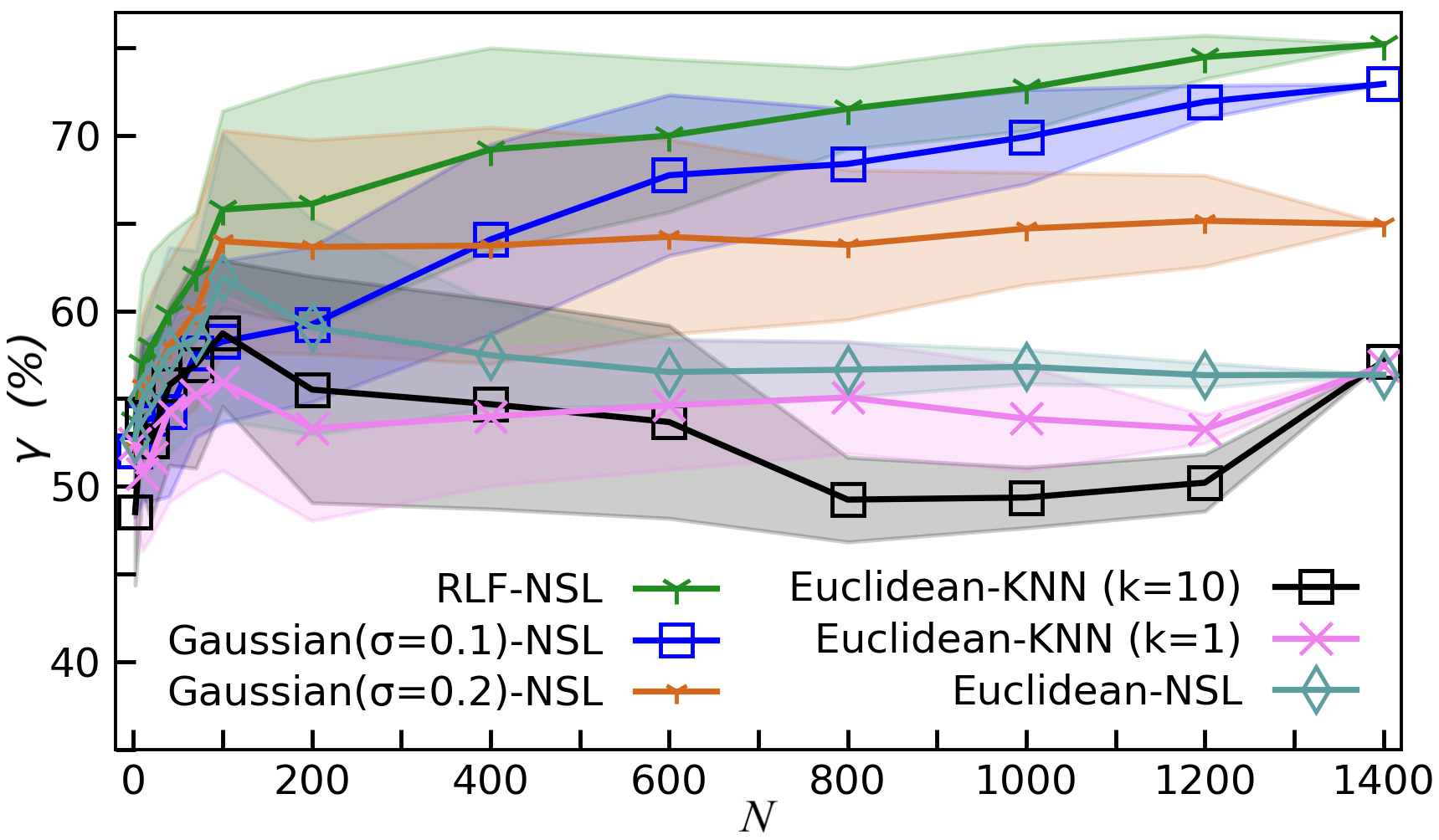}
	\caption{\R{(Color online) The testing accuracy of the RLF-NSL on the IMDb dataset comparing different kernels (Euclidean, Gaussian, and RLF) and classification strategies (KNN and NSL). The x-axis shows the number of labeled samples in each class.}}
	\label{fig-IMBD}
\end{figure}

\R{We compare the testing accuracy by using different kernels and classification strategies, as shown in Fig. \ref{fig-IMBD}. We choose IMDb \cite{imdbweb}, a recognized dataset in the field of natural language processing. Each sample is a comment on a movie, and the task is to predict whether it is positive or negative. The dataset contains 50000 samples, in which half for training and half for testing. For convenience, we limit the maximal number of the features in a sample (i.e., the maximal number of words in a comment) to $M_{\text{max}}=100$, and finally use 2773 training samples and 2963 testing samples. The labeled samples are randomly selected from the training samples, and the testing accuracy is evaluated by the testing samples. We test two classification strategies, which are KNN and NSL. We also compare different kernels. The Euclidean distance $f_E$ is given by Eq.~(\ref{eq-Euclidean}). For the Gaussian kernel, the distance is defined by a Gaussian distribution, which satisfies}
\begin{eqnarray}
	\R{f_G(\mathbf{x}^1, \mathbf{x}^2) = e^{-\frac{f^{2}_E(\mathbf{x}^1, \mathbf{x}^2)}{2 \sigma^2}}.}
	\label{eq-Gaussian}
\end{eqnarray}
\R{where $\sigma$ controls the variance. For the Euclidean-NSL algorithm, we use $f_E$ in Eq.~(\ref{eq-aveRLF}) to obtain the classifications. The rest parts are the same as RLF-NSL. For the Euclidean-KNN algorithm, we use $f_E$ to obtained the $k$ labeled samples with the smallest distances. In RLF-NSL, we flexibly adjust the rescaling factor $\beta$ as the number of labeled samples varies. The RLF-NSL achieves the highest testing accuracy among these algorithms.}

\begin{figure}[tbp]
	\includegraphics[width=1\linewidth]{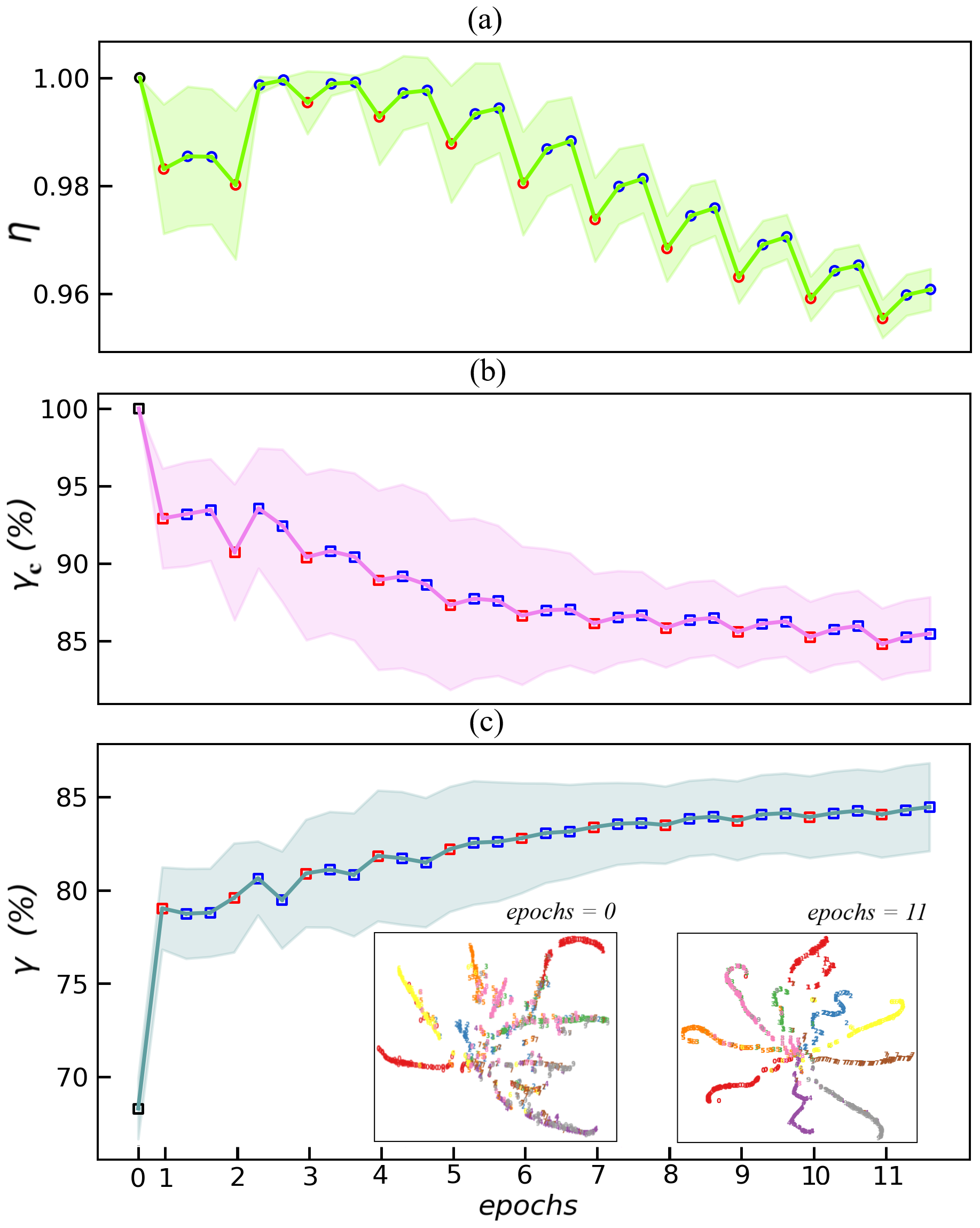}
	\caption{(Color online) For the RLF-NSSL with $N=6$ labeled samples in each class (few-shot case), (a) and (b) show the confidence $\eta$ and classification accuracy $\gamma_{c}$ of the samples in the clusters, respectively, in different epochs. (c) shows the classification accuracy $\gamma$ for the testing set. The insets of (c) illustrate the visualizations by applying t-SNE to the testing samples in the low-dimensional space [Eq. (\ref{eq-aveRLF})]. See the details in the main text.}
	\label{fig-confidence}
\end{figure}

To demonstrate how the classification precision is improved by updating the pseudo-labeled samples in the clusters, we take the few-shot learning by RLF-NSSL as an example. At the beginning, there are $N=6$ labeled samples in each class to define the cluster. For the zeroth epoch, the low-dimensional data and classifications are obtained by these labeled samples. In the update process, each epoch contains three sub-steps. In the first sub-step, we classify 500 samples with the highest RLF samples and add them to the corresponding clusters according to their classifications. In the second and third sub-steps, we update the clusters by replacing part of the pseudo-labeled samples. In specific, we move 500 samples that have the lowest confidence $\eta$ from each cluster to the pool of the unclassified samples. Then we calculate the classifications of all samples in the pool, and add the 500 samples with the highest RLF to the corresponding clusters.

In Fig. \ref{fig-confidence} (a) and (b), we show the confidence $\eta$ and classification accuracy $\gamma_{c}$ for the samples inside the clusters. Each time when we add new samples to the clusters in the first sub-step of each epoch (see red markers),  both $\eta$ and $\gamma_{c}$ decrease. By updating the clusters, we observe obvious improvement of $\eta$ by replacing the less confident samples in the clusters. Slight improvement of $\gamma_{c}$ is observed in general after the second sub-step. Even though the pseudo-labels of the samples in the clusters become less accurate as the clusters contain more and more pseudo samples, we observe monotonous increase (with insignificant fluctuations) of the testing accuracy $\gamma$ as shown in Fig. \ref{fig-confidence} (c). This is a convincing evidence on the validity of our RLF-NSSL and the pseudo-labeling strategy.

\section{Discussion from the perspective of rate reduction}

In Ref \cite{ma2007segmentation,yu2020learning}, several general principles were proposed on the continuous mapping from the original feature space to a low-dimensional space for the purposes of classification or clustering, known as the principles of maximal coding rate reduction (MCR$^{2}$). Considering the classification problems, the representations should satisfies the following properties: a) samples from the same class should belong to a low-dimensional linear subspaces; b) samples from different classes should belong to different low-dimensional subspaces and uncorrelated; c) the variance of features for the samples in the same class should be as large as possible as long as b) is satisfied. These three principles are known as \textit{within-class compressibility}, \textit{between-class discrimination}, and \textit{overall diversity}, respectively.

Our results imply that MCR$^{2}$ should also apply to the machine learning in the Hilbert space of many qubits. In our scheme, the clusters map each sample [the product state obtained by the feature map given by Eq. (\ref{eq-RLF})] to a $P$-dimensional vector. The distribution of these vectors is defined by their mutual distances based on the RLF. 

The insets of Fig. \ref{fig-confidence} (c) show the visualizations of the $P$-dimensional vectors from the testing set in at the 0th and 11th epochs. At the 0th epoch, we simply use the labeled samples to define the clusters. The testing accuracy is less than $70\%$. At the 11th epoch, the clusters consist of the labeled and pseudo-labeled samples. The pseudo-labeled samples are updated using the RLF-NSSL algorithm. The testing accuracy is round $85\%$. Comparing the two distributions in the two-dimensional space, it is obvious that at the 11th epoch, the samples in the same class incline to form a one-dimensional stream, indicating the within-class compressibility. The samples are distributed as ``radiations'' from the middle toward the edges, indicating the overall diversity. Each two neighboring radial lines give a similar angel, indicating the between-class discrimination. Inspecting from these three aspects, one can see that the samples at the 11th epoch better satisfy the MCR$^{2}$ than those at the 0th epoch. Similar phenomena can be observed in the insets of Fig. \ref{fig-base}. The $\beta$ giving higher testing accuracy would better obey the MCR$^{2}$, and \textit{vice versa}. These results provide preliminary evidence for the validity of MCR$^{2}$ for the machine learning in the Hilbert space.

\section{Summary}

In this work, we propose the rescaled logarithmic fidelity (RLF) to define a tunable quantum kernel function for characterizing the similarity between many-body states. The advantages of RLF are revealed by applying it for non-parametric semi-supervised learning. \R{Higher} accuracy is achieved particularly for the unsupervised and few-shot cases \R{compared with several well-established non-parametric methods}. \R{From the visualization of the data in the low-dimensional effective space defined by the RLF, our results support the applicability of MCR$^{2}$ theory. The proposed RLF and the semi-supervised learning strategies} can be readily applied to other QML methods, including those with parametric models (such as parametrized quantum circuits \cite{zhu2019training, benedetti2019generative,benedetti2019parameterized,chen2020variational,du2020expressive,cao2020cost,du2020expressive, xin2021experimental, cincio2021machine} and tensor networks \cite{socher2013reasoning,  stoudenmire2016supervised, han2018unsupervised, cheng2018information, chen2018equivalence, cheng2019tree, liu2019machine, huggins2019towards, schroder2019tensor, efthymiou2019tensornetwork, sun2020generative, sun2020tangent, guo2020tensor, ran2020tensor, cheng2021supervised, reyes2021multi}). Our code can be publicly found on Github \cite{code}.

\section*{Acknowledgment} The authors are thankful to Ding Liu and Zhang-Hao Zhou-Yin for stimulating discussions. This work was supported by NSFC (Grant No. 12004266 and No. 11834014), Beijing Natural Science Foundation (No. 1192005 and No. Z180013), Foundation of Beijing Education Committees (No. KM202010028013), and the key research project of Academy for Multidisciplinary Studies, Capital Normal University.

\bibliography{ref}

\begin{thebibliography}{74}%
\makeatletter
\providecommand \@ifxundefined [1]{%
 \@ifx{#1\undefined}
}%
\providecommand \@ifnum [1]{%
 \ifnum #1\expandafter \@firstoftwo
 \else \expandafter \@secondoftwo
 \fi
}%
\providecommand \@ifx [1]{%
 \ifx #1\expandafter \@firstoftwo
 \else \expandafter \@secondoftwo
 \fi
}%
\providecommand \natexlab [1]{#1}%
\providecommand \enquote  [1]{``#1''}%
\providecommand \bibnamefont  [1]{#1}%
\providecommand \bibfnamefont [1]{#1}%
\providecommand \citenamefont [1]{#1}%
\providecommand \href@noop [0]{\@secondoftwo}%
\providecommand \href [0]{\begingroup \@sanitize@url \@href}%
\providecommand \@href[1]{\@@startlink{#1}\@@href}%
\providecommand \@@href[1]{\endgroup#1\@@endlink}%
\providecommand \@sanitize@url [0]{\catcode `\\12\catcode `\$12\catcode
  `\&12\catcode `\#12\catcode `\^12\catcode `\_12\catcode `\%12\relax}%
\providecommand \@@startlink[1]{}%
\providecommand \@@endlink[0]{}%
\providecommand \url  [0]{\begingroup\@sanitize@url \@url }%
\providecommand \@url [1]{\endgroup\@href {#1}{\urlprefix }}%
\providecommand \urlprefix  [0]{URL }%
\providecommand \Eprint [0]{\href }%
\providecommand \doibase [0]{https://doi.org/}%
\providecommand \selectlanguage [0]{\@gobble}%
\providecommand \bibinfo  [0]{\@secondoftwo}%
\providecommand \bibfield  [0]{\@secondoftwo}%
\providecommand \translation [1]{[#1]}%
\providecommand \BibitemOpen [0]{}%
\providecommand \bibitemStop [0]{}%
\providecommand \bibitemNoStop [0]{.\EOS\space}%
\providecommand \EOS [0]{\spacefactor3000\relax}%
\providecommand \BibitemShut  [1]{\csname bibitem#1\endcsname}%
\let\auto@bib@innerbib\@empty
\bibitem [{\citenamefont {Shawe-Taylor}\ \emph {et~al.}(2004)\citenamefont
  {Shawe-Taylor}, \citenamefont {Cristianini} \emph
  {et~al.}}]{shawe2004kernel}%
  \BibitemOpen
  \bibfield  {author} {\bibinfo {author} {\bibfnamefont {J.}~\bibnamefont
  {Shawe-Taylor}}, \bibinfo {author} {\bibfnamefont {N.}~\bibnamefont
  {Cristianini}}, \emph {et~al.},\ }\href@noop {} {\emph {\bibinfo {title}
  {Kernel methods for pattern analysis}}}\ (\bibinfo  {publisher} {Cambridge
  university press},\ \bibinfo {year} {2004})\BibitemShut {NoStop}%
\bibitem [{\citenamefont {Hofmann}\ \emph {et~al.}(2008)\citenamefont
  {Hofmann}, \citenamefont {Schölkopf},\ and\ \citenamefont
  {Smola}}]{hofmann2008kernel}%
  \BibitemOpen
  \bibfield  {author} {\bibinfo {author} {\bibfnamefont {T.}~\bibnamefont
  {Hofmann}}, \bibinfo {author} {\bibfnamefont {B.}~\bibnamefont
  {Schölkopf}},\ and\ \bibinfo {author} {\bibfnamefont {A.~J.}\ \bibnamefont
  {Smola}},\ }\bibfield  {title} {\bibinfo {title} {{Kernel methods in machine
  learning}},\ }\href {https://doi.org/10.1214/009053607000000677} {\bibfield
  {journal} {\bibinfo  {journal} {The Annals of Statistics}\ }\textbf {\bibinfo
  {volume} {36}},\ \bibinfo {pages} {1171 } (\bibinfo {year}
  {2008})}\BibitemShut {NoStop}%
\bibitem [{\citenamefont {Biamonte}\ \emph {et~al.}(2017)\citenamefont
  {Biamonte}, \citenamefont {Wittek}, \citenamefont {Pancotti}, \citenamefont
  {Rebentrost}, \citenamefont {Wiebe},\ and\ \citenamefont
  {Lloyd}}]{biamonte2017quantum}%
  \BibitemOpen
  \bibfield  {author} {\bibinfo {author} {\bibfnamefont {J.}~\bibnamefont
  {Biamonte}}, \bibinfo {author} {\bibfnamefont {P.}~\bibnamefont {Wittek}},
  \bibinfo {author} {\bibfnamefont {N.}~\bibnamefont {Pancotti}}, \bibinfo
  {author} {\bibfnamefont {P.}~\bibnamefont {Rebentrost}}, \bibinfo {author}
  {\bibfnamefont {N.}~\bibnamefont {Wiebe}},\ and\ \bibinfo {author}
  {\bibfnamefont {S.}~\bibnamefont {Lloyd}},\ }\bibfield  {title} {\bibinfo
  {title} {Quantum machine learning},\ }\href
  {https://doi.org/10.1038/nature23474} {\bibfield  {journal} {\bibinfo
  {journal} {Nature}\ }\textbf {\bibinfo {volume} {549}},\ \bibinfo {pages}
  {195} (\bibinfo {year} {2017})}\BibitemShut {NoStop}%
\bibitem [{\citenamefont {Schuld}\ and\ \citenamefont
  {Killoran}(2019)}]{schuld2019quantum}%
  \BibitemOpen
  \bibfield  {author} {\bibinfo {author} {\bibfnamefont {M.}~\bibnamefont
  {Schuld}}\ and\ \bibinfo {author} {\bibfnamefont {N.}~\bibnamefont
  {Killoran}},\ }\bibfield  {title} {\bibinfo {title} {Quantum machine learning
  in feature hilbert spaces},\ }\href
  {https://doi.org/10.1103/PhysRevLett.122.040504} {\bibfield  {journal}
  {\bibinfo  {journal} {Phys. Rev. Lett.}\ }\textbf {\bibinfo {volume} {122}},\
  \bibinfo {pages} {040504} (\bibinfo {year} {2019})}\BibitemShut {NoStop}%
\bibitem [{\citenamefont {Havl{\'i}{\v{c}}ek}\ \emph
  {et~al.}(2019)\citenamefont {Havl{\'i}{\v{c}}ek}, \citenamefont
  {C{\'o}rcoles}, \citenamefont {Temme}, \citenamefont {Harrow}, \citenamefont
  {Kandala}, \citenamefont {Chow},\ and\ \citenamefont
  {Gambetta}}]{havlivcek2019supervised}%
  \BibitemOpen
  \bibfield  {author} {\bibinfo {author} {\bibfnamefont {V.}~\bibnamefont
  {Havl{\'i}{\v{c}}ek}}, \bibinfo {author} {\bibfnamefont {A.~D.}\ \bibnamefont
  {C{\'o}rcoles}}, \bibinfo {author} {\bibfnamefont {K.}~\bibnamefont {Temme}},
  \bibinfo {author} {\bibfnamefont {A.~W.}\ \bibnamefont {Harrow}}, \bibinfo
  {author} {\bibfnamefont {A.}~\bibnamefont {Kandala}}, \bibinfo {author}
  {\bibfnamefont {J.~M.}\ \bibnamefont {Chow}},\ and\ \bibinfo {author}
  {\bibfnamefont {J.~M.}\ \bibnamefont {Gambetta}},\ }\bibfield  {title}
  {\bibinfo {title} {Supervised learning with quantum-enhanced feature
  spaces},\ }\href {https://doi.org/10.1038/s41586-019-0980-2} {\bibfield
  {journal} {\bibinfo  {journal} {Nature}\ }\textbf {\bibinfo {volume} {567}},\
  \bibinfo {pages} {209} (\bibinfo {year} {2019})}\BibitemShut {NoStop}%
\bibitem [{\citenamefont {Lloyd}\ \emph {et~al.}(2020)\citenamefont {Lloyd},
  \citenamefont {Schuld}, \citenamefont {Ijaz}, \citenamefont {Izaac},\ and\
  \citenamefont {Killoran}}]{lloyd2020quantum}%
  \BibitemOpen
  \bibfield  {author} {\bibinfo {author} {\bibfnamefont {S.}~\bibnamefont
  {Lloyd}}, \bibinfo {author} {\bibfnamefont {M.}~\bibnamefont {Schuld}},
  \bibinfo {author} {\bibfnamefont {A.}~\bibnamefont {Ijaz}}, \bibinfo {author}
  {\bibfnamefont {J.}~\bibnamefont {Izaac}},\ and\ \bibinfo {author}
  {\bibfnamefont {N.}~\bibnamefont {Killoran}},\ }\href@noop {} {\bibinfo
  {title} {Quantum embeddings for machine learning}} (\bibinfo {year} {2020}),\
  \Eprint {https://arxiv.org/abs/2001.03622} {arXiv:2001.03622 [quant-ph]}
  \BibitemShut {NoStop}%
\bibitem [{\citenamefont {Schuld}(2021)}]{schuld2021supervised}%
  \BibitemOpen
  \bibfield  {author} {\bibinfo {author} {\bibfnamefont {M.}~\bibnamefont
  {Schuld}},\ }\href@noop {} {\bibinfo {title} {Supervised quantum machine
  learning models are kernel methods}} (\bibinfo {year} {2021}),\ \Eprint
  {https://arxiv.org/abs/2101.11020} {arXiv:2101.11020 [quant-ph]} \BibitemShut
  {NoStop}%
\bibitem [{\citenamefont {Wiebe}\ \emph {et~al.}(2012)\citenamefont {Wiebe},
  \citenamefont {Braun},\ and\ \citenamefont {Lloyd}}]{wiebe2012quantum}%
  \BibitemOpen
  \bibfield  {author} {\bibinfo {author} {\bibfnamefont {N.}~\bibnamefont
  {Wiebe}}, \bibinfo {author} {\bibfnamefont {D.}~\bibnamefont {Braun}},\ and\
  \bibinfo {author} {\bibfnamefont {S.}~\bibnamefont {Lloyd}},\ }\bibfield
  {title} {\bibinfo {title} {Quantum algorithm for data fitting},\ }\href
  {https://doi.org/10.1103/PhysRevLett.109.050505} {\bibfield  {journal}
  {\bibinfo  {journal} {Phys. Rev. Lett.}\ }\textbf {\bibinfo {volume} {109}},\
  \bibinfo {pages} {050505} (\bibinfo {year} {2012})}\BibitemShut {NoStop}%
\bibitem [{\citenamefont {Lloyd}\ \emph {et~al.}(2014)\citenamefont {Lloyd},
  \citenamefont {Mohseni},\ and\ \citenamefont
  {Rebentrost}}]{lloyd2014quantum}%
  \BibitemOpen
  \bibfield  {author} {\bibinfo {author} {\bibfnamefont {S.}~\bibnamefont
  {Lloyd}}, \bibinfo {author} {\bibfnamefont {M.}~\bibnamefont {Mohseni}},\
  and\ \bibinfo {author} {\bibfnamefont {P.}~\bibnamefont {Rebentrost}},\
  }\bibfield  {title} {\bibinfo {title} {Quantum principal component
  analysis},\ }\href {https://doi.org/10.1038/nphys3029} {\bibfield  {journal}
  {\bibinfo  {journal} {Nature Physics}\ }\textbf {\bibinfo {volume} {10}},\
  \bibinfo {pages} {631} (\bibinfo {year} {2014})}\BibitemShut {NoStop}%
\bibitem [{\citenamefont {Stoudenmire}\ and\ \citenamefont
  {Schwab}(2016)}]{stoudenmire2016supervised}%
  \BibitemOpen
  \bibfield  {author} {\bibinfo {author} {\bibfnamefont {E.}~\bibnamefont
  {Stoudenmire}}\ and\ \bibinfo {author} {\bibfnamefont {D.}~\bibnamefont
  {Schwab}},\ }\bibfield  {title} {\bibinfo {title} {Supervised learning with
  tensor networks},\ }\href@noop {} {\bibfield  {journal} {\bibinfo  {journal}
  {Advances in Neural Information Processing Systems}\ ,\ \bibinfo {pages}
  {4806}} (\bibinfo {year} {2016})}\BibitemShut {NoStop}%
\bibitem [{\citenamefont {Schuld}\ \emph {et~al.}(2016)\citenamefont {Schuld},
  \citenamefont {Sinayskiy},\ and\ \citenamefont
  {Petruccione}}]{schuld2016prediction}%
  \BibitemOpen
  \bibfield  {author} {\bibinfo {author} {\bibfnamefont {M.}~\bibnamefont
  {Schuld}}, \bibinfo {author} {\bibfnamefont {I.}~\bibnamefont {Sinayskiy}},\
  and\ \bibinfo {author} {\bibfnamefont {F.}~\bibnamefont {Petruccione}},\
  }\bibfield  {title} {\bibinfo {title} {Prediction by linear regression on a
  quantum computer},\ }\href {https://doi.org/10.1103/PhysRevA.94.022342}
  {\bibfield  {journal} {\bibinfo  {journal} {Phys. Rev. A}\ }\textbf {\bibinfo
  {volume} {94}},\ \bibinfo {pages} {022342} (\bibinfo {year}
  {2016})}\BibitemShut {NoStop}%
\bibitem [{\citenamefont {Benedetti}\ \emph {et~al.}(2017)\citenamefont
  {Benedetti}, \citenamefont {Realpe-G\'omez}, \citenamefont {Biswas},\ and\
  \citenamefont {Perdomo-Ortiz}}]{benedetti2017quantum}%
  \BibitemOpen
  \bibfield  {author} {\bibinfo {author} {\bibfnamefont {M.}~\bibnamefont
  {Benedetti}}, \bibinfo {author} {\bibfnamefont {J.}~\bibnamefont
  {Realpe-G\'omez}}, \bibinfo {author} {\bibfnamefont {R.}~\bibnamefont
  {Biswas}},\ and\ \bibinfo {author} {\bibfnamefont {A.}~\bibnamefont
  {Perdomo-Ortiz}},\ }\bibfield  {title} {\bibinfo {title} {Quantum-assisted
  learning of hardware-embedded probabilistic graphical models},\ }\href
  {https://doi.org/10.1103/PhysRevX.7.041052} {\bibfield  {journal} {\bibinfo
  {journal} {Phys. Rev. X}\ }\textbf {\bibinfo {volume} {7}},\ \bibinfo {pages}
  {041052} (\bibinfo {year} {2017})}\BibitemShut {NoStop}%
\bibitem [{\citenamefont {Schuld}\ \emph {et~al.}(2017)\citenamefont {Schuld},
  \citenamefont {Fingerhuth},\ and\ \citenamefont
  {Petruccione}}]{schuld2017implementing}%
  \BibitemOpen
  \bibfield  {author} {\bibinfo {author} {\bibfnamefont {M.}~\bibnamefont
  {Schuld}}, \bibinfo {author} {\bibfnamefont {M.}~\bibnamefont {Fingerhuth}},\
  and\ \bibinfo {author} {\bibfnamefont {F.}~\bibnamefont {Petruccione}},\
  }\bibfield  {title} {\bibinfo {title} {Implementing a distance-based
  classifier with a quantum interference circuit},\ }\href
  {https://doi.org/10.1209/0295-5075/119/60002} {\bibfield  {journal} {\bibinfo
   {journal} {{EPL} (Europhysics Letters)}\ }\textbf {\bibinfo {volume}
  {119}},\ \bibinfo {pages} {60002} (\bibinfo {year} {2017})}\BibitemShut
  {NoStop}%
\bibitem [{\citenamefont {Kerenidis}\ \emph {et~al.}(2019)\citenamefont
  {Kerenidis}, \citenamefont {Landman}, \citenamefont {Luongo},\ and\
  \citenamefont {Prakash}}]{kerenidis2019q}%
  \BibitemOpen
  \bibfield  {author} {\bibinfo {author} {\bibfnamefont {I.}~\bibnamefont
  {Kerenidis}}, \bibinfo {author} {\bibfnamefont {J.}~\bibnamefont {Landman}},
  \bibinfo {author} {\bibfnamefont {A.}~\bibnamefont {Luongo}},\ and\ \bibinfo
  {author} {\bibfnamefont {A.}~\bibnamefont {Prakash}},\ }\bibfield  {title}
  {\bibinfo {title} {{q-means: A quantum algorithm for unsupervised machine
  learning}},\ }in\ \href {https://hal.archives-ouvertes.fr/hal-02411662}
  {\emph {\bibinfo {booktitle} {{NeurIPS 2019}}}}\ (\bibinfo {address}
  {Vancouver, Canada},\ \bibinfo {year} {2019})\BibitemShut {NoStop}%
\bibitem [{\citenamefont {Zhao}\ \emph {et~al.}(2019)\citenamefont {Zhao},
  \citenamefont {Fitzsimons},\ and\ \citenamefont
  {Fitzsimons}}]{zhao2019quantum}%
  \BibitemOpen
  \bibfield  {author} {\bibinfo {author} {\bibfnamefont {Z.}~\bibnamefont
  {Zhao}}, \bibinfo {author} {\bibfnamefont {J.~K.}\ \bibnamefont
  {Fitzsimons}},\ and\ \bibinfo {author} {\bibfnamefont {J.~F.}\ \bibnamefont
  {Fitzsimons}},\ }\bibfield  {title} {\bibinfo {title} {Quantum-assisted
  gaussian process regression},\ }\href
  {https://doi.org/10.1103/PhysRevA.99.052331} {\bibfield  {journal} {\bibinfo
  {journal} {Phys. Rev. A}\ }\textbf {\bibinfo {volume} {99}},\ \bibinfo
  {pages} {052331} (\bibinfo {year} {2019})}\BibitemShut {NoStop}%
\bibitem [{\citenamefont {LaRose}\ and\ \citenamefont
  {Coyle}(2020)}]{larose2020robust}%
  \BibitemOpen
  \bibfield  {author} {\bibinfo {author} {\bibfnamefont {R.}~\bibnamefont
  {LaRose}}\ and\ \bibinfo {author} {\bibfnamefont {B.}~\bibnamefont {Coyle}},\
  }\bibfield  {title} {\bibinfo {title} {Robust data encodings for quantum
  classifiers},\ }\href {https://doi.org/10.1103/PhysRevA.102.032420}
  {\bibfield  {journal} {\bibinfo  {journal} {Phys. Rev. A}\ }\textbf {\bibinfo
  {volume} {102}},\ \bibinfo {pages} {032420} (\bibinfo {year}
  {2020})}\BibitemShut {NoStop}%
\bibitem [{\citenamefont {Huang}\ \emph {et~al.}(2021)\citenamefont {Huang},
  \citenamefont {Broughton}, \citenamefont {Mohseni}, \citenamefont {Babbush},
  \citenamefont {Boixo}, \citenamefont {Neven},\ and\ \citenamefont
  {McClean}}]{huang2021power}%
  \BibitemOpen
  \bibfield  {author} {\bibinfo {author} {\bibfnamefont {H.-Y.}\ \bibnamefont
  {Huang}}, \bibinfo {author} {\bibfnamefont {M.}~\bibnamefont {Broughton}},
  \bibinfo {author} {\bibfnamefont {M.}~\bibnamefont {Mohseni}}, \bibinfo
  {author} {\bibfnamefont {R.}~\bibnamefont {Babbush}}, \bibinfo {author}
  {\bibfnamefont {S.}~\bibnamefont {Boixo}}, \bibinfo {author} {\bibfnamefont
  {H.}~\bibnamefont {Neven}},\ and\ \bibinfo {author} {\bibfnamefont {J.~R.}\
  \bibnamefont {McClean}},\ }\bibfield  {title} {\bibinfo {title} {Power of
  data in quantum machine learning},\ }\href
  {https://doi.org/10.1038/s41467-021-22539-9} {\bibfield  {journal} {\bibinfo
  {journal} {Nature Communications}\ }\textbf {\bibinfo {volume} {12}},\
  \bibinfo {pages} {2631} (\bibinfo {year} {2021})}\BibitemShut {NoStop}%
\bibitem [{\citenamefont {Park}\ \emph {et~al.}(2020)\citenamefont {Park},
  \citenamefont {Blank},\ and\ \citenamefont {Petruccione}}]{park2020theory}%
  \BibitemOpen
  \bibfield  {author} {\bibinfo {author} {\bibfnamefont {D.~K.}\ \bibnamefont
  {Park}}, \bibinfo {author} {\bibfnamefont {C.}~\bibnamefont {Blank}},\ and\
  \bibinfo {author} {\bibfnamefont {F.}~\bibnamefont {Petruccione}},\
  }\bibfield  {title} {\bibinfo {title} {The theory of the quantum kernel-based
  binary classifier},\ }\href
  {https://doi.org/https://doi.org/10.1016/j.physleta.2020.126422} {\bibfield
  {journal} {\bibinfo  {journal} {Physics Letters A}\ }\textbf {\bibinfo
  {volume} {384}},\ \bibinfo {pages} {126422} (\bibinfo {year}
  {2020})}\BibitemShut {NoStop}%
\bibitem [{\citenamefont {Han}\ \emph {et~al.}(2018)\citenamefont {Han},
  \citenamefont {Wang}, \citenamefont {Fan}, \citenamefont {Wang},\ and\
  \citenamefont {Zhang}}]{han2018unsupervised}%
  \BibitemOpen
  \bibfield  {author} {\bibinfo {author} {\bibfnamefont {Z.-Y.}\ \bibnamefont
  {Han}}, \bibinfo {author} {\bibfnamefont {J.}~\bibnamefont {Wang}}, \bibinfo
  {author} {\bibfnamefont {H.}~\bibnamefont {Fan}}, \bibinfo {author}
  {\bibfnamefont {L.}~\bibnamefont {Wang}},\ and\ \bibinfo {author}
  {\bibfnamefont {P.}~\bibnamefont {Zhang}},\ }\bibfield  {title} {\bibinfo
  {title} {Unsupervised generative modeling using matrix product states},\
  }\href {https://doi.org/10.1103/PhysRevX.8.031012} {\bibfield  {journal}
  {\bibinfo  {journal} {Phys. Rev. X}\ }\textbf {\bibinfo {volume} {8}},\
  \bibinfo {pages} {031012} (\bibinfo {year} {2018})}\BibitemShut {NoStop}%
\bibitem [{\citenamefont {Liu}\ \emph {et~al.}(2019)\citenamefont {Liu},
  \citenamefont {Ran}, \citenamefont {Wittek}, \citenamefont {Peng},
  \citenamefont {Garc{\'{\i}}a}, \citenamefont {Su},\ and\ \citenamefont
  {Lewenstein}}]{liu2019machine}%
  \BibitemOpen
  \bibfield  {author} {\bibinfo {author} {\bibfnamefont {D.}~\bibnamefont
  {Liu}}, \bibinfo {author} {\bibfnamefont {S.-J.}\ \bibnamefont {Ran}},
  \bibinfo {author} {\bibfnamefont {P.}~\bibnamefont {Wittek}}, \bibinfo
  {author} {\bibfnamefont {C.}~\bibnamefont {Peng}}, \bibinfo {author}
  {\bibfnamefont {R.~B.}\ \bibnamefont {Garc{\'{\i}}a}}, \bibinfo {author}
  {\bibfnamefont {G.}~\bibnamefont {Su}},\ and\ \bibinfo {author}
  {\bibfnamefont {M.}~\bibnamefont {Lewenstein}},\ }\bibfield  {title}
  {\bibinfo {title} {Machine learning by unitary tensor network of hierarchical
  tree structure},\ }\href {https://doi.org/10.1088/1367-2630/ab31ef}
  {\bibfield  {journal} {\bibinfo  {journal} {New Journal of Physics}\ }\textbf
  {\bibinfo {volume} {21}},\ \bibinfo {pages} {073059} (\bibinfo {year}
  {2019})}\BibitemShut {NoStop}%
\bibitem [{\citenamefont {Sun}\ \emph {et~al.}(2020{\natexlab{a}})\citenamefont
  {Sun}, \citenamefont {Peng}, \citenamefont {Liu}, \citenamefont {Ran},\ and\
  \citenamefont {Su}}]{sun2020generative}%
  \BibitemOpen
  \bibfield  {author} {\bibinfo {author} {\bibfnamefont {Z.-Z.}\ \bibnamefont
  {Sun}}, \bibinfo {author} {\bibfnamefont {C.}~\bibnamefont {Peng}}, \bibinfo
  {author} {\bibfnamefont {D.}~\bibnamefont {Liu}}, \bibinfo {author}
  {\bibfnamefont {S.-J.}\ \bibnamefont {Ran}},\ and\ \bibinfo {author}
  {\bibfnamefont {G.}~\bibnamefont {Su}},\ }\bibfield  {title} {\bibinfo
  {title} {Generative tensor network classification model for supervised
  machine learning},\ }\href {https://doi.org/10.1103/PhysRevB.101.075135}
  {\bibfield  {journal} {\bibinfo  {journal} {Phys. Rev. B}\ }\textbf {\bibinfo
  {volume} {101}},\ \bibinfo {pages} {075135} (\bibinfo {year}
  {2020}{\natexlab{a}})}\BibitemShut {NoStop}%
\bibitem [{\citenamefont {Ran}\ \emph {et~al.}(2020)\citenamefont {Ran},
  \citenamefont {Sun}, \citenamefont {Fei}, \citenamefont {Su},\ and\
  \citenamefont {Lewenstein}}]{ran2020tensor}%
  \BibitemOpen
  \bibfield  {author} {\bibinfo {author} {\bibfnamefont {S.-J.}\ \bibnamefont
  {Ran}}, \bibinfo {author} {\bibfnamefont {Z.-Z.}\ \bibnamefont {Sun}},
  \bibinfo {author} {\bibfnamefont {S.-M.}\ \bibnamefont {Fei}}, \bibinfo
  {author} {\bibfnamefont {G.}~\bibnamefont {Su}},\ and\ \bibinfo {author}
  {\bibfnamefont {M.}~\bibnamefont {Lewenstein}},\ }\bibfield  {title}
  {\bibinfo {title} {Tensor network compressed sensing with unsupervised
  machine learning},\ }\href {https://doi.org/10.1103/PhysRevResearch.2.033293}
  {\bibfield  {journal} {\bibinfo  {journal} {Phys. Rev. Research}\ }\textbf
  {\bibinfo {volume} {2}},\ \bibinfo {pages} {033293} (\bibinfo {year}
  {2020})}\BibitemShut {NoStop}%
\bibitem [{\citenamefont {Wang}\ \emph {et~al.}(2020)\citenamefont {Wang},
  \citenamefont {Xiao}, \citenamefont {Yi}, \citenamefont {Ran},\ and\
  \citenamefont {Xue}}]{wang2020quantum}%
  \BibitemOpen
  \bibfield  {author} {\bibinfo {author} {\bibfnamefont {K.}~\bibnamefont
  {Wang}}, \bibinfo {author} {\bibfnamefont {L.}~\bibnamefont {Xiao}}, \bibinfo
  {author} {\bibfnamefont {W.}~\bibnamefont {Yi}}, \bibinfo {author}
  {\bibfnamefont {S.-J.}\ \bibnamefont {Ran}},\ and\ \bibinfo {author}
  {\bibfnamefont {P.}~\bibnamefont {Xue}},\ }\href@noop {} {\bibinfo {title}
  {Quantum image classifier with single photons}} (\bibinfo {year} {2020}),\
  \Eprint {https://arxiv.org/abs/2003.08551} {arXiv:2003.08551 [quant-ph]}
  \BibitemShut {NoStop}%
\bibitem [{\citenamefont {Nielsen}\ and\ \citenamefont
  {Chuang}(2002)}]{nielsen2002quantum}%
  \BibitemOpen
  \bibfield  {author} {\bibinfo {author} {\bibfnamefont {M.~A.}\ \bibnamefont
  {Nielsen}}\ and\ \bibinfo {author} {\bibfnamefont {I.}~\bibnamefont
  {Chuang}},\ }\href@noop {} {\bibinfo {title} {Quantum computation and quantum
  information}} (\bibinfo {year} {2002})\BibitemShut {NoStop}%
\bibitem [{\citenamefont {D'Ariano}\ and\ \citenamefont
  {Lo~Presti}(2001)}]{d2001quantum}%
  \BibitemOpen
  \bibfield  {author} {\bibinfo {author} {\bibfnamefont {G.~M.}\ \bibnamefont
  {D'Ariano}}\ and\ \bibinfo {author} {\bibfnamefont {P.}~\bibnamefont
  {Lo~Presti}},\ }\bibfield  {title} {\bibinfo {title} {Quantum tomography for
  measuring experimentally the matrix elements of an arbitrary quantum
  operation},\ }\href {https://doi.org/10.1103/PhysRevLett.86.4195} {\bibfield
  {journal} {\bibinfo  {journal} {Phys. Rev. Lett.}\ }\textbf {\bibinfo
  {volume} {86}},\ \bibinfo {pages} {4195} (\bibinfo {year}
  {2001})}\BibitemShut {NoStop}%
\bibitem [{\citenamefont {Buhrman}\ and\ \citenamefont
  {\v{S}palek}(2006)}]{buhrman2006quantum}%
  \BibitemOpen
  \bibfield  {author} {\bibinfo {author} {\bibfnamefont {H.}~\bibnamefont
  {Buhrman}}\ and\ \bibinfo {author} {\bibfnamefont {R.}~\bibnamefont
  {\v{S}palek}},\ }\bibfield  {title} {\bibinfo {title} {Quantum verification
  of matrix products},\ }in\ \href@noop {} {\emph {\bibinfo {booktitle}
  {Proceedings of the Seventeenth Annual ACM-SIAM Symposium on Discrete
  Algorithm}}},\ \bibinfo {series and number} {SODA '06}\ (\bibinfo
  {publisher} {Society for Industrial and Applied Mathematics},\ \bibinfo
  {address} {USA},\ \bibinfo {year} {2006})\ p.\ \bibinfo {pages}
  {880–889}\BibitemShut {NoStop}%
\bibitem [{\citenamefont {Zhou}\ \emph {et~al.}(2008)\citenamefont {Zhou},
  \citenamefont {Or\'us},\ and\ \citenamefont {Vidal}}]{zhou2008ground}%
  \BibitemOpen
  \bibfield  {author} {\bibinfo {author} {\bibfnamefont {H.-Q.}\ \bibnamefont
  {Zhou}}, \bibinfo {author} {\bibfnamefont {R.}~\bibnamefont {Or\'us}},\ and\
  \bibinfo {author} {\bibfnamefont {G.}~\bibnamefont {Vidal}},\ }\bibfield
  {title} {\bibinfo {title} {Ground state fidelity from tensor network
  representations},\ }\href {https://doi.org/10.1103/PhysRevLett.100.080601}
  {\bibfield  {journal} {\bibinfo  {journal} {Phys. Rev. Lett.}\ }\textbf
  {\bibinfo {volume} {100}},\ \bibinfo {pages} {080601} (\bibinfo {year}
  {2008})}\BibitemShut {NoStop}%
\bibitem [{\citenamefont {Abasto}\ \emph {et~al.}(2008)\citenamefont {Abasto},
  \citenamefont {Hamma},\ and\ \citenamefont {Zanardi}}]{abasto2008fidelity}%
  \BibitemOpen
  \bibfield  {author} {\bibinfo {author} {\bibfnamefont {D.~F.}\ \bibnamefont
  {Abasto}}, \bibinfo {author} {\bibfnamefont {A.}~\bibnamefont {Hamma}},\ and\
  \bibinfo {author} {\bibfnamefont {P.}~\bibnamefont {Zanardi}},\ }\bibfield
  {title} {\bibinfo {title} {Fidelity analysis of topological quantum phase
  transitions},\ }\href {https://doi.org/10.1103/PhysRevA.78.010301} {\bibfield
   {journal} {\bibinfo  {journal} {Phys. Rev. A}\ }\textbf {\bibinfo {volume}
  {78}},\ \bibinfo {pages} {010301} (\bibinfo {year} {2008})}\BibitemShut
  {NoStop}%
\bibitem [{\citenamefont {Schwandt}\ \emph {et~al.}(2009)\citenamefont
  {Schwandt}, \citenamefont {Alet},\ and\ \citenamefont
  {Capponi}}]{schwandt2009quantum}%
  \BibitemOpen
  \bibfield  {author} {\bibinfo {author} {\bibfnamefont {D.}~\bibnamefont
  {Schwandt}}, \bibinfo {author} {\bibfnamefont {F.}~\bibnamefont {Alet}},\
  and\ \bibinfo {author} {\bibfnamefont {S.}~\bibnamefont {Capponi}},\
  }\bibfield  {title} {\bibinfo {title} {Quantum monte carlo simulations of
  fidelity at magnetic quantum phase transitions},\ }\href
  {https://doi.org/10.1103/PhysRevLett.103.170501} {\bibfield  {journal}
  {\bibinfo  {journal} {Phys. Rev. Lett.}\ }\textbf {\bibinfo {volume} {103}},\
  \bibinfo {pages} {170501} (\bibinfo {year} {2009})}\BibitemShut {NoStop}%
\bibitem [{\citenamefont {Quan}\ and\ \citenamefont
  {Cucchietti}(2009)}]{quan2009quantum}%
  \BibitemOpen
  \bibfield  {author} {\bibinfo {author} {\bibfnamefont {H.~T.}\ \bibnamefont
  {Quan}}\ and\ \bibinfo {author} {\bibfnamefont {F.~M.}\ \bibnamefont
  {Cucchietti}},\ }\bibfield  {title} {\bibinfo {title} {Quantum fidelity and
  thermal phase transitions},\ }\href
  {https://doi.org/10.1103/PhysRevE.79.031101} {\bibfield  {journal} {\bibinfo
  {journal} {Phys. Rev. E}\ }\textbf {\bibinfo {volume} {79}},\ \bibinfo
  {pages} {031101} (\bibinfo {year} {2009})}\BibitemShut {NoStop}%
\bibitem [{\citenamefont {Zhao}\ and\ \citenamefont
  {Zhou}(2009)}]{zhao2009singularities}%
  \BibitemOpen
  \bibfield  {author} {\bibinfo {author} {\bibfnamefont {J.-H.}\ \bibnamefont
  {Zhao}}\ and\ \bibinfo {author} {\bibfnamefont {H.-Q.}\ \bibnamefont
  {Zhou}},\ }\bibfield  {title} {\bibinfo {title} {Singularities in
  ground-state fidelity and quantum phase transitions for the kitaev model},\
  }\href {https://doi.org/10.1103/PhysRevB.80.014403} {\bibfield  {journal}
  {\bibinfo  {journal} {Phys. Rev. B}\ }\textbf {\bibinfo {volume} {80}},\
  \bibinfo {pages} {014403} (\bibinfo {year} {2009})}\BibitemShut {NoStop}%
\bibitem [{\citenamefont {Xiong}\ \emph {et~al.}(2009)\citenamefont {Xiong},
  \citenamefont {Ma}, \citenamefont {Sun},\ and\ \citenamefont
  {Wang}}]{xiong2009reduced}%
  \BibitemOpen
  \bibfield  {author} {\bibinfo {author} {\bibfnamefont {H.-N.}\ \bibnamefont
  {Xiong}}, \bibinfo {author} {\bibfnamefont {J.}~\bibnamefont {Ma}}, \bibinfo
  {author} {\bibfnamefont {Z.}~\bibnamefont {Sun}},\ and\ \bibinfo {author}
  {\bibfnamefont {X.}~\bibnamefont {Wang}},\ }\bibfield  {title} {\bibinfo
  {title} {Reduced-fidelity approach for quantum phase transitions in
  spin-$\frac{1}{2}$ dimerized heisenberg chains},\ }\href
  {https://doi.org/10.1103/PhysRevB.79.174425} {\bibfield  {journal} {\bibinfo
  {journal} {Phys. Rev. B}\ }\textbf {\bibinfo {volume} {79}},\ \bibinfo
  {pages} {174425} (\bibinfo {year} {2009})}\BibitemShut {NoStop}%
\bibitem [{\citenamefont {Ran}(2020)}]{ran2020encoding}%
  \BibitemOpen
  \bibfield  {author} {\bibinfo {author} {\bibfnamefont {S.-J.}\ \bibnamefont
  {Ran}},\ }\bibfield  {title} {\bibinfo {title} {Encoding of matrix product
  states into quantum circuits of one- and two-qubit gates},\ }\href
  {https://doi.org/10.1103/PhysRevA.101.032310} {\bibfield  {journal} {\bibinfo
   {journal} {Phys. Rev. A}\ }\textbf {\bibinfo {volume} {101}},\ \bibinfo
  {pages} {032310} (\bibinfo {year} {2020})}\BibitemShut {NoStop}%
\bibitem [{\citenamefont {Yang}\ \emph {et~al.}(2021)\citenamefont {Yang},
  \citenamefont {Sun}, \citenamefont {Ran},\ and\ \citenamefont
  {Su}}]{yang2021visualizing}%
  \BibitemOpen
  \bibfield  {author} {\bibinfo {author} {\bibfnamefont {Y.}~\bibnamefont
  {Yang}}, \bibinfo {author} {\bibfnamefont {Z.-Z.}\ \bibnamefont {Sun}},
  \bibinfo {author} {\bibfnamefont {S.-J.}\ \bibnamefont {Ran}},\ and\ \bibinfo
  {author} {\bibfnamefont {G.}~\bibnamefont {Su}},\ }\bibfield  {title}
  {\bibinfo {title} {Visualizing quantum phases and identifying quantum phase
  transitions by nonlinear dimensional reduction},\ }\href
  {https://doi.org/10.1103/PhysRevB.103.075106} {\bibfield  {journal} {\bibinfo
   {journal} {Phys. Rev. B}\ }\textbf {\bibinfo {volume} {103}},\ \bibinfo
  {pages} {075106} (\bibinfo {year} {2021})}\BibitemShut {NoStop}%
\bibitem [{\citenamefont {Van~der Maaten}\ and\ \citenamefont
  {Hinton}(2008)}]{van2008visualizing}%
  \BibitemOpen
  \bibfield  {author} {\bibinfo {author} {\bibfnamefont {L.}~\bibnamefont
  {Van~der Maaten}}\ and\ \bibinfo {author} {\bibfnamefont {G.}~\bibnamefont
  {Hinton}},\ }\bibfield  {title} {\bibinfo {title} {Visualizing data using
  t-sne.},\ }\href@noop {} {\bibfield  {journal} {\bibinfo  {journal} {Journal
  of machine learning research}\ }\textbf {\bibinfo {volume} {9}} (\bibinfo
  {year} {2008})}\BibitemShut {NoStop}%
\bibitem [{\citenamefont {Ma}\ \emph {et~al.}(2007)\citenamefont {Ma},
  \citenamefont {Derksen}, \citenamefont {Hong},\ and\ \citenamefont
  {Wright}}]{ma2007segmentation}%
  \BibitemOpen
  \bibfield  {author} {\bibinfo {author} {\bibfnamefont {Y.}~\bibnamefont
  {Ma}}, \bibinfo {author} {\bibfnamefont {H.}~\bibnamefont {Derksen}},
  \bibinfo {author} {\bibfnamefont {W.}~\bibnamefont {Hong}},\ and\ \bibinfo
  {author} {\bibfnamefont {J.}~\bibnamefont {Wright}},\ }\bibfield  {title}
  {\bibinfo {title} {Segmentation of multivariate mixed data via lossy data
  coding and compression},\ }\href {https://doi.org/10.1109/TPAMI.2007.1085}
  {\bibfield  {journal} {\bibinfo  {journal} {IEEE Transactions on Pattern
  Analysis and Machine Intelligence}\ }\textbf {\bibinfo {volume} {29}},\
  \bibinfo {pages} {1546} (\bibinfo {year} {2007})}\BibitemShut {NoStop}%
\bibitem [{\citenamefont {Yu}\ \emph {et~al.}()\citenamefont {Yu},
  \citenamefont {Chan}, \citenamefont {You}, \citenamefont {Song},\ and\
  \citenamefont {Ma}}]{yu2020learning}%
  \BibitemOpen
  \bibfield  {author} {\bibinfo {author} {\bibfnamefont {Y.}~\bibnamefont
  {Yu}}, \bibinfo {author} {\bibfnamefont {K.~H.}\ \bibnamefont {Chan}},
  \bibinfo {author} {\bibfnamefont {C.}~\bibnamefont {You}}, \bibinfo {author}
  {\bibfnamefont {C.}~\bibnamefont {Song}},\ and\ \bibinfo {author}
  {\bibfnamefont {Y.}~\bibnamefont {Ma}},\ }\bibfield  {title} {\bibinfo
  {title} {Learning diverse and discriminative representations via the
  principle of maximal coding rate reduction},\ }\href
  {https://par.nsf.gov/biblio/10250387} {\bibinfo  {journal} {Advances in
  neural information processing systems}\ }\BibitemShut {NoStop}%
\bibitem [{\citenamefont {Socher}\ \emph {et~al.}(2013)\citenamefont {Socher},
  \citenamefont {Chen}, \citenamefont {Manning},\ and\ \citenamefont
  {Ng}}]{socher2013reasoning}%
  \BibitemOpen
\bibfield  {journal} {  }\bibfield  {author} {\bibinfo {author} {\bibfnamefont
  {R.}~\bibnamefont {Socher}}, \bibinfo {author} {\bibfnamefont
  {D.}~\bibnamefont {Chen}}, \bibinfo {author} {\bibfnamefont {C.~D.}\
  \bibnamefont {Manning}},\ and\ \bibinfo {author} {\bibfnamefont
  {A.}~\bibnamefont {Ng}},\ }\bibfield  {title} {\bibinfo {title} {Reasoning
  with neural tensor networks for knowledge base completion},\ }in\ \href
  {https://proceedings.neurips.cc/paper/2013/file/b337e84de8752b27eda3a12363109e80-Paper.pdf}
  {\emph {\bibinfo {booktitle} {Advances in Neural Information Processing
  Systems}}},\ Vol.~\bibinfo {volume} {26},\ \bibinfo {editor} {edited by\
  \bibinfo {editor} {\bibfnamefont {C.~J.~C.}\ \bibnamefont {Burges}}, \bibinfo
  {editor} {\bibfnamefont {L.}~\bibnamefont {Bottou}}, \bibinfo {editor}
  {\bibfnamefont {M.}~\bibnamefont {Welling}}, \bibinfo {editor} {\bibfnamefont
  {Z.}~\bibnamefont {Ghahramani}},\ and\ \bibinfo {editor} {\bibfnamefont
  {K.~Q.}\ \bibnamefont {Weinberger}}}\ (\bibinfo  {publisher} {Curran
  Associates, Inc.},\ \bibinfo {year} {2013})\BibitemShut {NoStop}%
\bibitem [{\citenamefont {Cheng}\ \emph {et~al.}(2018)\citenamefont {Cheng},
  \citenamefont {Chen},\ and\ \citenamefont {Wang}}]{cheng2018information}%
  \BibitemOpen
  \bibfield  {author} {\bibinfo {author} {\bibfnamefont {S.}~\bibnamefont
  {Cheng}}, \bibinfo {author} {\bibfnamefont {J.}~\bibnamefont {Chen}},\ and\
  \bibinfo {author} {\bibfnamefont {L.}~\bibnamefont {Wang}},\ }\bibfield
  {title} {\bibinfo {title} {Information perspective to probabilistic modeling:
  Boltzmann machines versus born machines},\ }\bibfield  {journal} {\bibinfo
  {journal} {Entropy}\ }\textbf {\bibinfo {volume} {20}},\ \href
  {https://doi.org/10.3390/e20080583} {10.3390/e20080583} (\bibinfo {year}
  {2018})\BibitemShut {NoStop}%
\bibitem [{\citenamefont {Chen}\ \emph {et~al.}(2018)\citenamefont {Chen},
  \citenamefont {Cheng}, \citenamefont {Xie}, \citenamefont {Wang},\ and\
  \citenamefont {Xiang}}]{chen2018equivalence}%
  \BibitemOpen
  \bibfield  {author} {\bibinfo {author} {\bibfnamefont {J.}~\bibnamefont
  {Chen}}, \bibinfo {author} {\bibfnamefont {S.}~\bibnamefont {Cheng}},
  \bibinfo {author} {\bibfnamefont {H.}~\bibnamefont {Xie}}, \bibinfo {author}
  {\bibfnamefont {L.}~\bibnamefont {Wang}},\ and\ \bibinfo {author}
  {\bibfnamefont {T.}~\bibnamefont {Xiang}},\ }\bibfield  {title} {\bibinfo
  {title} {Equivalence of restricted boltzmann machines and tensor network
  states},\ }\href {https://doi.org/10.1103/PhysRevB.97.085104} {\bibfield
  {journal} {\bibinfo  {journal} {Phys. Rev. B}\ }\textbf {\bibinfo {volume}
  {97}},\ \bibinfo {pages} {085104} (\bibinfo {year} {2018})}\BibitemShut
  {NoStop}%
\bibitem [{\citenamefont {Cheng}\ \emph {et~al.}(2019)\citenamefont {Cheng},
  \citenamefont {Wang}, \citenamefont {Xiang},\ and\ \citenamefont
  {Zhang}}]{cheng2019tree}%
  \BibitemOpen
  \bibfield  {author} {\bibinfo {author} {\bibfnamefont {S.}~\bibnamefont
  {Cheng}}, \bibinfo {author} {\bibfnamefont {L.}~\bibnamefont {Wang}},
  \bibinfo {author} {\bibfnamefont {T.}~\bibnamefont {Xiang}},\ and\ \bibinfo
  {author} {\bibfnamefont {P.}~\bibnamefont {Zhang}},\ }\bibfield  {title}
  {\bibinfo {title} {Tree tensor networks for generative modeling},\ }\href
  {https://doi.org/10.1103/PhysRevB.99.155131} {\bibfield  {journal} {\bibinfo
  {journal} {Phys. Rev. B}\ }\textbf {\bibinfo {volume} {99}},\ \bibinfo
  {pages} {155131} (\bibinfo {year} {2019})}\BibitemShut {NoStop}%
\bibitem [{\citenamefont {Huggins}\ \emph {et~al.}(2019)\citenamefont
  {Huggins}, \citenamefont {Patil}, \citenamefont {Mitchell}, \citenamefont
  {Whaley},\ and\ \citenamefont {Stoudenmire}}]{huggins2019towards}%
  \BibitemOpen
  \bibfield  {author} {\bibinfo {author} {\bibfnamefont {W.}~\bibnamefont
  {Huggins}}, \bibinfo {author} {\bibfnamefont {P.}~\bibnamefont {Patil}},
  \bibinfo {author} {\bibfnamefont {B.}~\bibnamefont {Mitchell}}, \bibinfo
  {author} {\bibfnamefont {K.~B.}\ \bibnamefont {Whaley}},\ and\ \bibinfo
  {author} {\bibfnamefont {E.~M.}\ \bibnamefont {Stoudenmire}},\ }\bibfield
  {title} {\bibinfo {title} {Towards quantum machine learning with tensor
  networks},\ }\href {https://doi.org/10.1088/2058-9565/aaea94} {\bibfield
  {journal} {\bibinfo  {journal} {Quantum Science and Technology}\ }\textbf
  {\bibinfo {volume} {4}},\ \bibinfo {pages} {024001} (\bibinfo {year}
  {2019})}\BibitemShut {NoStop}%
\bibitem [{\citenamefont {Schr{\"o}der}\ \emph {et~al.}(2019)\citenamefont
  {Schr{\"o}der}, \citenamefont {Turban}, \citenamefont {Musser}, \citenamefont
  {Hine},\ and\ \citenamefont {Chin}}]{schroder2019tensor}%
  \BibitemOpen
  \bibfield  {author} {\bibinfo {author} {\bibfnamefont {F.~A. Y.~N.}\
  \bibnamefont {Schr{\"o}der}}, \bibinfo {author} {\bibfnamefont {D.~H.~P.}\
  \bibnamefont {Turban}}, \bibinfo {author} {\bibfnamefont {A.~J.}\
  \bibnamefont {Musser}}, \bibinfo {author} {\bibfnamefont {N.~D.~M.}\
  \bibnamefont {Hine}},\ and\ \bibinfo {author} {\bibfnamefont {A.~W.}\
  \bibnamefont {Chin}},\ }\bibfield  {title} {\bibinfo {title} {Tensor network
  simulation of multi-environmental open quantum dynamics via machine learning
  and entanglement renormalisation},\ }\href
  {https://doi.org/10.1038/s41467-019-09039-7} {\bibfield  {journal} {\bibinfo
  {journal} {Nature Communications}\ }\textbf {\bibinfo {volume} {10}},\
  \bibinfo {pages} {1062} (\bibinfo {year} {2019})}\BibitemShut {NoStop}%
\bibitem [{\citenamefont {Efthymiou}\ \emph {et~al.}(2019)\citenamefont
  {Efthymiou}, \citenamefont {Hidary},\ and\ \citenamefont
  {Leichenauer}}]{efthymiou2019tensornetwork}%
  \BibitemOpen
  \bibfield  {author} {\bibinfo {author} {\bibfnamefont {S.}~\bibnamefont
  {Efthymiou}}, \bibinfo {author} {\bibfnamefont {J.}~\bibnamefont {Hidary}},\
  and\ \bibinfo {author} {\bibfnamefont {S.}~\bibnamefont {Leichenauer}},\
  }\href@noop {} {\bibinfo {title} {Tensornetwork for machine learning}}
  (\bibinfo {year} {2019}),\ \Eprint {https://arxiv.org/abs/1906.06329}
  {arXiv:1906.06329 [cs.LG]} \BibitemShut {NoStop}%
\bibitem [{\citenamefont {Sun}\ \emph {et~al.}(2020{\natexlab{b}})\citenamefont
  {Sun}, \citenamefont {Ran},\ and\ \citenamefont {Su}}]{sun2020tangent}%
  \BibitemOpen
  \bibfield  {author} {\bibinfo {author} {\bibfnamefont {Z.-Z.}\ \bibnamefont
  {Sun}}, \bibinfo {author} {\bibfnamefont {S.-J.}\ \bibnamefont {Ran}},\ and\
  \bibinfo {author} {\bibfnamefont {G.}~\bibnamefont {Su}},\ }\bibfield
  {title} {\bibinfo {title} {Tangent-space gradient optimization of tensor
  network for machine learning},\ }\href
  {https://doi.org/10.1103/PhysRevE.102.012152} {\bibfield  {journal} {\bibinfo
   {journal} {Phys. Rev. E}\ }\textbf {\bibinfo {volume} {102}},\ \bibinfo
  {pages} {012152} (\bibinfo {year} {2020}{\natexlab{b}})}\BibitemShut
  {NoStop}%
\bibitem [{\citenamefont {Guo}\ \emph {et~al.}(2020)\citenamefont {Guo},
  \citenamefont {Modi},\ and\ \citenamefont {Poletti}}]{guo2020tensor}%
  \BibitemOpen
  \bibfield  {author} {\bibinfo {author} {\bibfnamefont {C.}~\bibnamefont
  {Guo}}, \bibinfo {author} {\bibfnamefont {K.}~\bibnamefont {Modi}},\ and\
  \bibinfo {author} {\bibfnamefont {D.}~\bibnamefont {Poletti}},\ }\bibfield
  {title} {\bibinfo {title} {Tensor-network-based machine learning of
  non-markovian quantum processes},\ }\href
  {https://doi.org/10.1103/PhysRevA.102.062414} {\bibfield  {journal} {\bibinfo
   {journal} {Phys. Rev. A}\ }\textbf {\bibinfo {volume} {102}},\ \bibinfo
  {pages} {062414} (\bibinfo {year} {2020})}\BibitemShut {NoStop}%
\bibitem [{\citenamefont {Cheng}\ \emph {et~al.}(2021)\citenamefont {Cheng},
  \citenamefont {Wang},\ and\ \citenamefont {Zhang}}]{cheng2021supervised}%
  \BibitemOpen
  \bibfield  {author} {\bibinfo {author} {\bibfnamefont {S.}~\bibnamefont
  {Cheng}}, \bibinfo {author} {\bibfnamefont {L.}~\bibnamefont {Wang}},\ and\
  \bibinfo {author} {\bibfnamefont {P.}~\bibnamefont {Zhang}},\ }\bibfield
  {title} {\bibinfo {title} {Supervised learning with projected entangled pair
  states},\ }\href {https://doi.org/10.1103/PhysRevB.103.125117} {\bibfield
  {journal} {\bibinfo  {journal} {Phys. Rev. B}\ }\textbf {\bibinfo {volume}
  {103}},\ \bibinfo {pages} {125117} (\bibinfo {year} {2021})}\BibitemShut
  {NoStop}%
\bibitem [{\citenamefont {Reyes}\ and\ \citenamefont
  {Stoudenmire}(2021)}]{reyes2021multi}%
  \BibitemOpen
  \bibfield  {author} {\bibinfo {author} {\bibfnamefont {J.}~\bibnamefont
  {Reyes}}\ and\ \bibinfo {author} {\bibfnamefont {E.~M.}\ \bibnamefont
  {Stoudenmire}},\ }\bibfield  {title} {\bibinfo {title} {A multi-scale tensor
  network architecture for machine learning},\ }\href
  {http://iopscience.iop.org/article/10.1088/2632-2153/abffe8} {\bibfield
  {journal} {\bibinfo  {journal} {Machine Learning: Science and Technology}\ }
  (\bibinfo {year} {2021})}\BibitemShut {NoStop}%
\bibitem [{\citenamefont {Zhu}\ \emph {et~al.}(2019)\citenamefont {Zhu},
  \citenamefont {Linke}, \citenamefont {Benedetti}, \citenamefont {Landsman},
  \citenamefont {Nguyen}, \citenamefont {Alderete}, \citenamefont
  {Perdomo-Ortiz}, \citenamefont {Korda}, \citenamefont {Garfoot},
  \citenamefont {Brecque}, \citenamefont {Egan}, \citenamefont {Perdomo},\ and\
  \citenamefont {Monroe}}]{zhu2019training}%
  \BibitemOpen
  \bibfield  {author} {\bibinfo {author} {\bibfnamefont {D.}~\bibnamefont
  {Zhu}}, \bibinfo {author} {\bibfnamefont {N.~M.}\ \bibnamefont {Linke}},
  \bibinfo {author} {\bibfnamefont {M.}~\bibnamefont {Benedetti}}, \bibinfo
  {author} {\bibfnamefont {K.~A.}\ \bibnamefont {Landsman}}, \bibinfo {author}
  {\bibfnamefont {N.~H.}\ \bibnamefont {Nguyen}}, \bibinfo {author}
  {\bibfnamefont {C.~H.}\ \bibnamefont {Alderete}}, \bibinfo {author}
  {\bibfnamefont {A.}~\bibnamefont {Perdomo-Ortiz}}, \bibinfo {author}
  {\bibfnamefont {N.}~\bibnamefont {Korda}}, \bibinfo {author} {\bibfnamefont
  {A.}~\bibnamefont {Garfoot}}, \bibinfo {author} {\bibfnamefont
  {C.}~\bibnamefont {Brecque}}, \bibinfo {author} {\bibfnamefont
  {L.}~\bibnamefont {Egan}}, \bibinfo {author} {\bibfnamefont {O.}~\bibnamefont
  {Perdomo}},\ and\ \bibinfo {author} {\bibfnamefont {C.}~\bibnamefont
  {Monroe}},\ }\bibfield  {title} {\bibinfo {title} {Training of quantum
  circuits on a hybrid quantum computer},\ }\bibfield  {journal} {\bibinfo
  {journal} {Science Advances}\ }\textbf {\bibinfo {volume} {5}},\ \href
  {https://doi.org/10.1126/sciadv.aaw9918} {10.1126/sciadv.aaw9918} (\bibinfo
  {year} {2019}),\ \Eprint
  {https://arxiv.org/abs/https://advances.sciencemag.org/content/5/10/eaaw9918.full.pdf}
  {https://advances.sciencemag.org/content/5/10/eaaw9918.full.pdf} \BibitemShut
  {NoStop}%
\bibitem [{\citenamefont {Benedetti}\ \emph
  {et~al.}(2019{\natexlab{a}})\citenamefont {Benedetti}, \citenamefont
  {Garcia-Pintos}, \citenamefont {Perdomo}, \citenamefont {Leyton-Ortega},
  \citenamefont {Nam},\ and\ \citenamefont
  {Perdomo-Ortiz}}]{benedetti2019generative}%
  \BibitemOpen
  \bibfield  {author} {\bibinfo {author} {\bibfnamefont {M.}~\bibnamefont
  {Benedetti}}, \bibinfo {author} {\bibfnamefont {D.}~\bibnamefont
  {Garcia-Pintos}}, \bibinfo {author} {\bibfnamefont {O.}~\bibnamefont
  {Perdomo}}, \bibinfo {author} {\bibfnamefont {V.}~\bibnamefont
  {Leyton-Ortega}}, \bibinfo {author} {\bibfnamefont {Y.}~\bibnamefont {Nam}},\
  and\ \bibinfo {author} {\bibfnamefont {A.}~\bibnamefont {Perdomo-Ortiz}},\
  }\bibfield  {title} {\bibinfo {title} {A generative modeling approach for
  benchmarking and training shallow quantum circuits},\ }\href
  {https://doi.org/10.1038/s41534-019-0157-8} {\bibfield  {journal} {\bibinfo
  {journal} {npj Quantum Information}\ }\textbf {\bibinfo {volume} {5}},\
  \bibinfo {pages} {45} (\bibinfo {year} {2019}{\natexlab{a}})}\BibitemShut
  {NoStop}%
\bibitem [{\citenamefont {Benedetti}\ \emph
  {et~al.}(2019{\natexlab{b}})\citenamefont {Benedetti}, \citenamefont {Lloyd},
  \citenamefont {Sack},\ and\ \citenamefont
  {Fiorentini}}]{benedetti2019parameterized}%
  \BibitemOpen
  \bibfield  {author} {\bibinfo {author} {\bibfnamefont {M.}~\bibnamefont
  {Benedetti}}, \bibinfo {author} {\bibfnamefont {E.}~\bibnamefont {Lloyd}},
  \bibinfo {author} {\bibfnamefont {S.}~\bibnamefont {Sack}},\ and\ \bibinfo
  {author} {\bibfnamefont {M.}~\bibnamefont {Fiorentini}},\ }\bibfield  {title}
  {\bibinfo {title} {Parameterized quantum circuits as machine learning
  models},\ }\href {https://doi.org/10.1088/2058-9565/ab4eb5} {\bibfield
  {journal} {\bibinfo  {journal} {Quantum Science and Technology}\ }\textbf
  {\bibinfo {volume} {4}},\ \bibinfo {pages} {043001} (\bibinfo {year}
  {2019}{\natexlab{b}})}\BibitemShut {NoStop}%
\bibitem [{\citenamefont {Chen}\ \emph {et~al.}(2020)\citenamefont {Chen},
  \citenamefont {Yang}, \citenamefont {Qi}, \citenamefont {Chen}, \citenamefont
  {Ma},\ and\ \citenamefont {Goan}}]{chen2020variational}%
  \BibitemOpen
  \bibfield  {author} {\bibinfo {author} {\bibfnamefont {S.~Y.-C.}\
  \bibnamefont {Chen}}, \bibinfo {author} {\bibfnamefont {C.-H.~H.}\
  \bibnamefont {Yang}}, \bibinfo {author} {\bibfnamefont {J.}~\bibnamefont
  {Qi}}, \bibinfo {author} {\bibfnamefont {P.-Y.}\ \bibnamefont {Chen}},
  \bibinfo {author} {\bibfnamefont {X.}~\bibnamefont {Ma}},\ and\ \bibinfo
  {author} {\bibfnamefont {H.-S.}\ \bibnamefont {Goan}},\ }\bibfield  {title}
  {\bibinfo {title} {Variational quantum circuits for deep reinforcement
  learning},\ }\href {https://doi.org/10.1109/ACCESS.2020.3010470} {\bibfield
  {journal} {\bibinfo  {journal} {IEEE Access}\ }\textbf {\bibinfo {volume}
  {8}},\ \bibinfo {pages} {141007} (\bibinfo {year} {2020})}\BibitemShut
  {NoStop}%
\bibitem [{\citenamefont {Du}\ \emph {et~al.}(2020)\citenamefont {Du},
  \citenamefont {Hsieh}, \citenamefont {Liu},\ and\ \citenamefont
  {Tao}}]{du2020expressive}%
  \BibitemOpen
  \bibfield  {author} {\bibinfo {author} {\bibfnamefont {Y.}~\bibnamefont
  {Du}}, \bibinfo {author} {\bibfnamefont {M.-H.}\ \bibnamefont {Hsieh}},
  \bibinfo {author} {\bibfnamefont {T.}~\bibnamefont {Liu}},\ and\ \bibinfo
  {author} {\bibfnamefont {D.}~\bibnamefont {Tao}},\ }\bibfield  {title}
  {\bibinfo {title} {Expressive power of parametrized quantum circuits},\
  }\href {https://doi.org/10.1103/PhysRevResearch.2.033125} {\bibfield
  {journal} {\bibinfo  {journal} {Phys. Rev. Research}\ }\textbf {\bibinfo
  {volume} {2}},\ \bibinfo {pages} {033125} (\bibinfo {year}
  {2020})}\BibitemShut {NoStop}%
\bibitem [{\citenamefont {Cao}\ \emph {et~al.}(2020)\citenamefont {Cao},
  \citenamefont {Wossnig}, \citenamefont {Vlastakis}, \citenamefont {Leek},\
  and\ \citenamefont {Grant}}]{cao2020cost}%
  \BibitemOpen
  \bibfield  {author} {\bibinfo {author} {\bibfnamefont {S.}~\bibnamefont
  {Cao}}, \bibinfo {author} {\bibfnamefont {L.}~\bibnamefont {Wossnig}},
  \bibinfo {author} {\bibfnamefont {B.}~\bibnamefont {Vlastakis}}, \bibinfo
  {author} {\bibfnamefont {P.}~\bibnamefont {Leek}},\ and\ \bibinfo {author}
  {\bibfnamefont {E.}~\bibnamefont {Grant}},\ }\bibfield  {title} {\bibinfo
  {title} {Cost-function embedding and dataset encoding for machine learning
  with parametrized quantum circuits},\ }\href
  {https://doi.org/10.1103/PhysRevA.101.052309} {\bibfield  {journal} {\bibinfo
   {journal} {Phys. Rev. A}\ }\textbf {\bibinfo {volume} {101}},\ \bibinfo
  {pages} {052309} (\bibinfo {year} {2020})}\BibitemShut {NoStop}%
\bibitem [{\citenamefont {Xin}\ \emph {et~al.}(2021)\citenamefont {Xin},
  \citenamefont {Che}, \citenamefont {Xi}, \citenamefont {Singh}, \citenamefont
  {Nie}, \citenamefont {Li}, \citenamefont {Dong},\ and\ \citenamefont
  {Lu}}]{xin2021experimental}%
  \BibitemOpen
  \bibfield  {author} {\bibinfo {author} {\bibfnamefont {T.}~\bibnamefont
  {Xin}}, \bibinfo {author} {\bibfnamefont {L.}~\bibnamefont {Che}}, \bibinfo
  {author} {\bibfnamefont {C.}~\bibnamefont {Xi}}, \bibinfo {author}
  {\bibfnamefont {A.}~\bibnamefont {Singh}}, \bibinfo {author} {\bibfnamefont
  {X.}~\bibnamefont {Nie}}, \bibinfo {author} {\bibfnamefont {J.}~\bibnamefont
  {Li}}, \bibinfo {author} {\bibfnamefont {Y.}~\bibnamefont {Dong}},\ and\
  \bibinfo {author} {\bibfnamefont {D.}~\bibnamefont {Lu}},\ }\bibfield
  {title} {\bibinfo {title} {Experimental quantum principal component analysis
  via parametrized quantum circuits},\ }\href
  {https://doi.org/10.1103/PhysRevLett.126.110502} {\bibfield  {journal}
  {\bibinfo  {journal} {Phys. Rev. Lett.}\ }\textbf {\bibinfo {volume} {126}},\
  \bibinfo {pages} {110502} (\bibinfo {year} {2021})}\BibitemShut {NoStop}%
\bibitem [{\citenamefont {Cincio}\ \emph {et~al.}(2021)\citenamefont {Cincio},
  \citenamefont {Rudinger}, \citenamefont {Sarovar},\ and\ \citenamefont
  {Coles}}]{cincio2021machine}%
  \BibitemOpen
  \bibfield  {author} {\bibinfo {author} {\bibfnamefont {L.}~\bibnamefont
  {Cincio}}, \bibinfo {author} {\bibfnamefont {K.}~\bibnamefont {Rudinger}},
  \bibinfo {author} {\bibfnamefont {M.}~\bibnamefont {Sarovar}},\ and\ \bibinfo
  {author} {\bibfnamefont {P.~J.}\ \bibnamefont {Coles}},\ }\bibfield  {title}
  {\bibinfo {title} {Machine learning of noise-resilient quantum circuits},\
  }\href {https://doi.org/10.1103/PRXQuantum.2.010324} {\bibfield  {journal}
  {\bibinfo  {journal} {PRX Quantum}\ }\textbf {\bibinfo {volume} {2}},\
  \bibinfo {pages} {010324} (\bibinfo {year} {2021})}\BibitemShut {NoStop}%
\bibitem [{\citenamefont {Farhi}\ and\ \citenamefont
  {Neven}(2018)}]{farhi2018classification}%
  \BibitemOpen
  \bibfield  {author} {\bibinfo {author} {\bibfnamefont {E.}~\bibnamefont
  {Farhi}}\ and\ \bibinfo {author} {\bibfnamefont {H.}~\bibnamefont {Neven}},\
  }\href@noop {} {\bibinfo {title} {Classification with quantum neural networks
  on near term processors}} (\bibinfo {year} {2018}),\ \Eprint
  {https://arxiv.org/abs/1802.06002} {arXiv:1802.06002 [quant-ph]} \BibitemShut
  {NoStop}%
\bibitem [{\citenamefont {McClean}\ \emph {et~al.}(2018)\citenamefont
  {McClean}, \citenamefont {Boixo}, \citenamefont {Smelyanskiy}, \citenamefont
  {Babbush},\ and\ \citenamefont {Neven}}]{mcclean2018barren}%
  \BibitemOpen
  \bibfield  {author} {\bibinfo {author} {\bibfnamefont {J.~R.}\ \bibnamefont
  {McClean}}, \bibinfo {author} {\bibfnamefont {S.}~\bibnamefont {Boixo}},
  \bibinfo {author} {\bibfnamefont {V.~N.}\ \bibnamefont {Smelyanskiy}},
  \bibinfo {author} {\bibfnamefont {R.}~\bibnamefont {Babbush}},\ and\ \bibinfo
  {author} {\bibfnamefont {H.}~\bibnamefont {Neven}},\ }\bibfield  {title}
  {\bibinfo {title} {Barren plateaus in quantum neural network training
  landscapes},\ }\href {https://doi.org/10.1038/s41467-018-07090-4} {\bibfield
  {journal} {\bibinfo  {journal} {Nature Communications}\ }\textbf {\bibinfo
  {volume} {9}},\ \bibinfo {pages} {4812} (\bibinfo {year} {2018})}\BibitemShut
  {NoStop}%
\bibitem [{\citenamefont {Cong}\ \emph {et~al.}(2019)\citenamefont {Cong},
  \citenamefont {Choi},\ and\ \citenamefont {Lukin}}]{cong2019quantum}%
  \BibitemOpen
  \bibfield  {author} {\bibinfo {author} {\bibfnamefont {I.}~\bibnamefont
  {Cong}}, \bibinfo {author} {\bibfnamefont {S.}~\bibnamefont {Choi}},\ and\
  \bibinfo {author} {\bibfnamefont {M.~D.}\ \bibnamefont {Lukin}},\ }\bibfield
  {title} {\bibinfo {title} {Quantum convolutional neural networks},\ }\href
  {https://doi.org/10.1038/s41567-019-0648-8} {\bibfield  {journal} {\bibinfo
  {journal} {Nature Physics}\ }\textbf {\bibinfo {volume} {15}},\ \bibinfo
  {pages} {1273} (\bibinfo {year} {2019})}\BibitemShut {NoStop}%
\bibitem [{\citenamefont {Killoran}\ \emph {et~al.}(2019)\citenamefont
  {Killoran}, \citenamefont {Bromley}, \citenamefont {Arrazola}, \citenamefont
  {Schuld}, \citenamefont {Quesada},\ and\ \citenamefont
  {Lloyd}}]{killoran2019continuous}%
  \BibitemOpen
  \bibfield  {author} {\bibinfo {author} {\bibfnamefont {N.}~\bibnamefont
  {Killoran}}, \bibinfo {author} {\bibfnamefont {T.~R.}\ \bibnamefont
  {Bromley}}, \bibinfo {author} {\bibfnamefont {J.~M.}\ \bibnamefont
  {Arrazola}}, \bibinfo {author} {\bibfnamefont {M.}~\bibnamefont {Schuld}},
  \bibinfo {author} {\bibfnamefont {N.}~\bibnamefont {Quesada}},\ and\ \bibinfo
  {author} {\bibfnamefont {S.}~\bibnamefont {Lloyd}},\ }\bibfield  {title}
  {\bibinfo {title} {Continuous-variable quantum neural networks},\ }\href
  {https://doi.org/10.1103/PhysRevResearch.1.033063} {\bibfield  {journal}
  {\bibinfo  {journal} {Phys. Rev. Research}\ }\textbf {\bibinfo {volume}
  {1}},\ \bibinfo {pages} {033063} (\bibinfo {year} {2019})}\BibitemShut
  {NoStop}%
\bibitem [{\citenamefont {Mari}\ \emph {et~al.}(2020)\citenamefont {Mari},
  \citenamefont {Bromley}, \citenamefont {Izaac}, \citenamefont {Schuld},\ and\
  \citenamefont {Killoran}}]{mari2020transfer}%
  \BibitemOpen
  \bibfield  {author} {\bibinfo {author} {\bibfnamefont {A.}~\bibnamefont
  {Mari}}, \bibinfo {author} {\bibfnamefont {T.~R.}\ \bibnamefont {Bromley}},
  \bibinfo {author} {\bibfnamefont {J.}~\bibnamefont {Izaac}}, \bibinfo
  {author} {\bibfnamefont {M.}~\bibnamefont {Schuld}},\ and\ \bibinfo {author}
  {\bibfnamefont {N.}~\bibnamefont {Killoran}},\ }\bibfield  {title} {\bibinfo
  {title} {Transfer learning in hybrid classical-quantum neural networks},\
  }\href {https://doi.org/10.22331/q-2020-10-09-340} {\bibfield  {journal}
  {\bibinfo  {journal} {{Quantum}}\ }\textbf {\bibinfo {volume} {4}},\ \bibinfo
  {pages} {340} (\bibinfo {year} {2020})}\BibitemShut {NoStop}%
\bibitem [{\citenamefont {Beer}\ \emph {et~al.}(2020)\citenamefont {Beer},
  \citenamefont {Bondarenko}, \citenamefont {Farrelly}, \citenamefont
  {Osborne}, \citenamefont {Salzmann}, \citenamefont {Scheiermann},\ and\
  \citenamefont {Wolf}}]{beer2020training}%
  \BibitemOpen
  \bibfield  {author} {\bibinfo {author} {\bibfnamefont {K.}~\bibnamefont
  {Beer}}, \bibinfo {author} {\bibfnamefont {D.}~\bibnamefont {Bondarenko}},
  \bibinfo {author} {\bibfnamefont {T.}~\bibnamefont {Farrelly}}, \bibinfo
  {author} {\bibfnamefont {T.~J.}\ \bibnamefont {Osborne}}, \bibinfo {author}
  {\bibfnamefont {R.}~\bibnamefont {Salzmann}}, \bibinfo {author}
  {\bibfnamefont {D.}~\bibnamefont {Scheiermann}},\ and\ \bibinfo {author}
  {\bibfnamefont {R.}~\bibnamefont {Wolf}},\ }\bibfield  {title} {\bibinfo
  {title} {Training deep quantum neural networks},\ }\href
  {https://doi.org/10.1038/s41467-020-14454-2} {\bibfield  {journal} {\bibinfo
  {journal} {Nature Communications}\ }\textbf {\bibinfo {volume} {11}},\
  \bibinfo {pages} {808} (\bibinfo {year} {2020})}\BibitemShut {NoStop}%
\bibitem [{\citenamefont {Shen}\ \emph {et~al.}(2020)\citenamefont {Shen},
  \citenamefont {Zhang}, \citenamefont {You},\ and\ \citenamefont
  {Zhai}}]{shen2020information}%
  \BibitemOpen
  \bibfield  {author} {\bibinfo {author} {\bibfnamefont {H.}~\bibnamefont
  {Shen}}, \bibinfo {author} {\bibfnamefont {P.}~\bibnamefont {Zhang}},
  \bibinfo {author} {\bibfnamefont {Y.-Z.}\ \bibnamefont {You}},\ and\ \bibinfo
  {author} {\bibfnamefont {H.}~\bibnamefont {Zhai}},\ }\bibfield  {title}
  {\bibinfo {title} {Information scrambling in quantum neural networks},\
  }\href {https://doi.org/10.1103/PhysRevLett.124.200504} {\bibfield  {journal}
  {\bibinfo  {journal} {Phys. Rev. Lett.}\ }\textbf {\bibinfo {volume} {124}},\
  \bibinfo {pages} {200504} (\bibinfo {year} {2020})}\BibitemShut {NoStop}%
\bibitem [{mni()}]{mnistweb}%
  \BibitemOpen
  \href@noop {} {}\bibinfo {howpublished} {The website of MNIST dataset
  \url{http://yann.lecun.com/exdb/mnist/ }}\BibitemShut {NoStop}%
\bibitem [{\citenamefont {Cover}\ and\ \citenamefont
  {Hart}(1967)}]{cover1967nearest}%
  \BibitemOpen
  \bibfield  {author} {\bibinfo {author} {\bibfnamefont {T.}~\bibnamefont
  {Cover}}\ and\ \bibinfo {author} {\bibfnamefont {P.}~\bibnamefont {Hart}},\
  }\bibfield  {title} {\bibinfo {title} {Nearest neighbor pattern
  classification},\ }\href {https://doi.org/10.1109/TIT.1967.1053964}
  {\bibfield  {journal} {\bibinfo  {journal} {IEEE Transactions on Information
  Theory}\ }\textbf {\bibinfo {volume} {13}},\ \bibinfo {pages} {21} (\bibinfo
  {year} {1967})}\BibitemShut {NoStop}%
\bibitem [{\citenamefont {Langley}\ \emph {et~al.}(1992)\citenamefont
  {Langley}, \citenamefont {Iba}, \citenamefont {Thompson} \emph
  {et~al.}}]{langley1992analysis}%
  \BibitemOpen
  \bibfield  {author} {\bibinfo {author} {\bibfnamefont {P.}~\bibnamefont
  {Langley}}, \bibinfo {author} {\bibfnamefont {W.}~\bibnamefont {Iba}},
  \bibinfo {author} {\bibfnamefont {K.}~\bibnamefont {Thompson}}, \emph
  {et~al.},\ }\bibfield  {title} {\bibinfo {title} {An analysis of bayesian
  classifiers},\ }in\ \href@noop {} {\emph {\bibinfo {booktitle} {Aaai}}},\
  Vol.~\bibinfo {volume} {90}\ (\bibinfo {organization} {Citeseer},\ \bibinfo
  {year} {1992})\ pp.\ \bibinfo {pages} {223--228}\BibitemShut {NoStop}%
\bibitem [{\citenamefont {Munkres}(1957)}]{munkres1957algorithms}%
  \BibitemOpen
  \bibfield  {author} {\bibinfo {author} {\bibfnamefont {J.}~\bibnamefont
  {Munkres}},\ }\bibfield  {title} {\bibinfo {title} {Algorithms for the
  assignment and transportation problems},\ }\href
  {https://doi.org/10.1137/0105003} {\bibfield  {journal} {\bibinfo  {journal}
  {Journal of the Society for Industrial and Applied Mathematics}\ }\textbf
  {\bibinfo {volume} {5}},\ \bibinfo {pages} {32} (\bibinfo {year} {1957})},\
  \Eprint {https://arxiv.org/abs/https://doi.org/10.1137/0105003}
  {https://doi.org/10.1137/0105003} \BibitemShut {NoStop}%
\bibitem [{\citenamefont {Sarle}(1990)}]{sarle1990algorithms}%
  \BibitemOpen
  \bibfield  {author} {\bibinfo {author} {\bibfnamefont {W.~S.}\ \bibnamefont
  {Sarle}},\ }\href@noop {} {\bibinfo {title} {Algorithms for clustering data}}
  (\bibinfo {year} {1990})\BibitemShut {NoStop}%
\bibitem [{\citenamefont {Kaufman}\ and\ \citenamefont
  {Rousseeuw}(2009)}]{kaufman2009finding}%
  \BibitemOpen
  \bibfield  {author} {\bibinfo {author} {\bibfnamefont {L.}~\bibnamefont
  {Kaufman}}\ and\ \bibinfo {author} {\bibfnamefont {P.~J.}\ \bibnamefont
  {Rousseeuw}},\ }\href@noop {} {\emph {\bibinfo {title} {Finding groups in
  data: an introduction to cluster analysis}}},\ Vol.\ \bibinfo {volume} {344}\
  (\bibinfo  {publisher} {John Wiley \& Sons},\ \bibinfo {year}
  {2009})\BibitemShut {NoStop}%
\bibitem [{\citenamefont {Mehrotra}\ \emph {et~al.}(1997)\citenamefont
  {Mehrotra}, \citenamefont {Mohan},\ and\ \citenamefont
  {Ranka}}]{mehrotra1997elements}%
  \BibitemOpen
  \bibfield  {author} {\bibinfo {author} {\bibfnamefont {K.}~\bibnamefont
  {Mehrotra}}, \bibinfo {author} {\bibfnamefont {C.~K.}\ \bibnamefont
  {Mohan}},\ and\ \bibinfo {author} {\bibfnamefont {S.}~\bibnamefont {Ranka}},\
  }\href@noop {} {\emph {\bibinfo {title} {Elements of artificial neural
  networks}}}\ (\bibinfo  {publisher} {MIT press},\ \bibinfo {year}
  {1997})\BibitemShut {NoStop}%
\bibitem [{\citenamefont {Ng}\ \emph {et~al.}(2001)\citenamefont {Ng},
  \citenamefont {Jordan},\ and\ \citenamefont {Weiss}}]{ng2001spectral}%
  \BibitemOpen
  \bibfield  {author} {\bibinfo {author} {\bibfnamefont {A.~Y.}\ \bibnamefont
  {Ng}}, \bibinfo {author} {\bibfnamefont {M.~I.}\ \bibnamefont {Jordan}},\
  and\ \bibinfo {author} {\bibfnamefont {Y.}~\bibnamefont {Weiss}},\ }\bibfield
   {title} {\bibinfo {title} {On spectral clustering: Analysis and an
  algorithm}\ }(\bibinfo  {publisher} {MIT Press},\ \bibinfo {address}
  {Cambridge, MA, USA},\ \bibinfo {year} {2001})\ p.\ \bibinfo {pages}
  {849–856}\BibitemShut {NoStop}%
\bibitem [{\citenamefont {Kamvar}\ \emph {et~al.}(2003)\citenamefont {Kamvar},
  \citenamefont {Sepandar}, \citenamefont {Klein}, \citenamefont {Dan},
  \citenamefont {Manning},\ and\ \citenamefont
  {Christopher}}]{kamvar2003spectral}%
  \BibitemOpen
  \bibfield  {author} {\bibinfo {author} {\bibfnamefont {K.}~\bibnamefont
  {Kamvar}}, \bibinfo {author} {\bibfnamefont {S.}~\bibnamefont {Sepandar}},
  \bibinfo {author} {\bibfnamefont {K.}~\bibnamefont {Klein}}, \bibinfo
  {author} {\bibfnamefont {D.}~\bibnamefont {Dan}}, \bibinfo {author}
  {\bibfnamefont {M.}~\bibnamefont {Manning}},\ and\ \bibinfo {author}
  {\bibfnamefont {C.}~\bibnamefont {Christopher}},\ }\href
  {http://ilpubs.stanford.edu:8090/587/} {\emph {\bibinfo {title} {Spectral
  Learning}}},\ \bibinfo {type} {Technical Report}\ \bibinfo {number}
  {2003-25}\ (\bibinfo  {institution} {Stanford InfoLab},\ \bibinfo {year}
  {2003})\BibitemShut {NoStop}%
\bibitem [{imd()}]{imdbweb}%
  \BibitemOpen
  \href@noop {} {}\bibinfo {howpublished} {The website of IMDb dataset
  \url{http://ai.stanford.edu/~amaas/data/sentiment/ }}\BibitemShut {NoStop}%
\bibitem [{cod()}]{code}%
  \BibitemOpen
  \href@noop {} {}\bibinfo {howpublished}
  {\url{https://github.com/Li-Wei-Ming/rlf.git }}\BibitemShut {NoStop}%
\end{thebibliography}%





\end{document}